# Insights and alternative proposals on the phase retrieval in high resolution transmission electron microscopy

Usha Bhat, and Ranjan Datta[1,2,*]

[1,2]*International Center for Materials Science, Chemistry and Physics of Materials Unit, JNCASR, Bangalore 560064.*

## Abstract

Alternative reconstruction method is proposed on retrieving the object exit wave function (OEW) directly from the recorded image intensity pattern in high resolution transmission electron microscopy (HRTEM). The method is based on applying a modified intensity equation representing the HRTEM image. A comparative discussion is provided between the existing methodologies involved in reconstruction of OEW, off-axis electron holography and the present proposal. Phase shift extracted from the experimental images of $MoS_2$, BN and ZnO are found to be in excellent agreement with the theoretical reference values. Additionally, it is shown that the Fourier series expansion of diffraction pattern is effective in retrieving the isolated and periodic image functions of certain form directly. However, for aperiodic object information e.g., defects, dopants, edges etc., the first method works in entirety.

*Corresponding author e-mail: ranjan@jncasr.ac.in

## 1. Introduction

Phase ($\phi$) is the fundamental quantity in quantitative high-resolution transmission electron interferometry [1–6]. The change in phase ($\Delta\phi$) of the probe electron wave after interacting with the object potential leads to the formation of specific intensity patterns in the respective image and diffraction planes. The intensity patterns can be recorded through any suitable imaging device, e.g., a charge-coupled device (CCD) camera. In case of transmission electron diffraction, $\Delta\phi$ carry information not only on the crystallographic phase of the scattering potential along high symmetry orientation but also on the electrostatic potential from the atoms and crystals that is essential for the identification and counting of atoms, extracting information on the chemical bonding from the experimental images [7–13]. There are few existing experimental and associated numerical phase retrieval methods in high resolution transmission electron microscopy e.g., through focal image series reconstruction based on HRTEM [2,14,15], atomic resolution off-axis electron holography [16,17], fitting object function directly by intensive computer simulation or so-called direct method [9], transport of intensity equation (TIE) [18], and phase velocity [7]. Some of the techniques mentioned above work both at medium and atomic resolution and TIE method was developed for the medium resolution applications.

Among various methods, complexities involved in object exit wave (OEW) reconstruction based on conventional through focal HRTEM image series is addressed in the present discussion and its analogy and differences with respect to off-axis electron holography are highlighted. Defocus HRTEM image is equivalent to in-line holography. In off-axis electron holography, the wave interference occurs at an angle between the reference and the object waves. The retrieval of OEW is performed at first by Fourier transformation (FT) of the image containing electron interference pattern, then selecting one of the two side bands (SBs) which are complex conjugate (or twin image) to each other followed by inverse-FT [4]. This procedure isolates the central band (CB) and the twin image wavefunctions from the recorded image. The phase and amplitude can be evaluated either by the arctan function corresponding to inverse-FT or fitting the inverse-FT pattern with the help of image simulation. As the starting data is the image, therefore, FT procedure does not lead to loss of any information in terms of crystallographic phase and inverse FT can return the image intensity pattern. Deconvolution of coherent aberration envelope can be performed posteriori that modifies the aberration figure in the image plane.

In case of HRTEM, the CB and twin image wave functions overlap in the diffraction plane, and FT procedure cannot separate them in the frequency space unlike off-axis electron holography [19,20]. Therefore, almost all the reconstruction methods in HRTEM involves multiplying the image intensity pattern recorded at different focus settings with a complex filter function consisting of coherent aberration envelope corresponding to each focus and then summing up over all the images to eliminate the unwanted twin image and non-linear components from the wanted OEW function. The complexities involved with reducing the unwanted components from the wanted OEW function led to the development of several reconstruction algorithms [15].

However, in the present manuscript it is demonstrated that by marginally modifying the form of the intensity equation describing the HRTEM image and concomitant justification, it is possible to retrieve the phase information directly from the atomic resolution images. The same equation in an intermediate form based on the wavefunction formalism is used in the existing OEW reconstruction procedures both in the case of in-line and off-axis electron holography. Two types of phases are of importance here. One is crystallographic phase describing the scattering distribution of potential corresponding to isolated or periodic arrangement of atoms forming specific image or diffraction pattern in the respective planes. The second one is the change in phase of probe electron wave due to strength of atomic potential, which is equivalent to electron density in X-ray crystallography, that determines the intensity of the dots in the image pattern. To understand the workings of the modified equation and underlying complexities involved in the existing methods, at first the flow of phase information from the object to the Gaussian image and diffraction planes are briefly discussed based on the various existing formalism i.e. Fresnel Huygens construction, weak phase object approximation (WPOA), transmitted wave function involving atom scattering factor derived from the Schrödinger integral equation and Fourier method similar to Abbe's approach of image formation [21–23]. Then, elemental principles behind existing OEW reconstruction methods are described. The phase change and associated retrieval of OEW function due to object, lens imperfections and geometry of interference are dependent on the ways the change in phase is incorporated and the form of the equations used in the mathematical formulations. However, it is shown that the phase information both in terms of crystallographic phase at atomic resolution and object potential are not lost in the high-resolution image intensity pattern for both in-line and off-axis electron holography where a reference wave is present. The presence of reference wave ensures the modulation of the intensity pattern and retaining the phase information in the image plane. The results obtained for $MoS_2$, BN and ZnO are found to be in excellent agreement with the theoretical reference values within the specified resolution limits. The working of the method is in accordance with the Born rule of probability amplitude and addresses the twin image issue in in-line holography. However, both the phase information is completely lost for the intensity pattern recorded in the diffraction plane. But it is shown that it is also possible to retrieve the complete phase information directly from the diffracted intensity alone. This latter method is based on the cosine based Fourier series expansion of the diffraction pattern that is similar to the zero phase retrieval based on Patterson function in X-ray crystallography for small molecular systems [24,25]. This latter method works exactly for certain type of functions both in isolated and periodic form. However, for aperiodic object information e.g., defects and dopants the first method works in its entirety.

## 2. Experimental Techniques

TEM samples of $MoS_2$ and BN layered materials were prepared by ultrasonication of respective powders (Sigma Aldrich) for 40 mins to exfoliate monolayers and few layers followed by drop casting on a holey carbon grid [26]. Cross sectional TEM specimen of ZnO epitaxial thin film was prepared by first mechanical thinning and then Ar ion polishing to perforation. HRTEM images were recorded in an aberration corrected FEI TITAN$^{3TM}$ 80-300 kV transmission electron microscope operating at 300 kV with optimum settings of $C_S$ and defocus [27,28]. ZnO epitaxial thin film was grown homoepitaxially on a (0001) ZnO substrate by pulsed laser deposition (PLD) following a specific growth procedure as describe in Ref.[6,29].

## 3. Brief overview on various conceptual elements of image formation, flow of phase information and OEW reconstruction

In the following sub-sections, at first various fundamental concepts and associated mathematical formalism corresponding to the flow of phase information from the object plane to the image and diffraction planes are described. Subsequently, a discussion is provided on the existing reconstruction methods based on HRTEM through focal image series before proceeding for the alternative proposals on OEW reconstruction in sec. 4.

### A. Methods based on Fresnel-Huygens construction and Fourier transformation

A typical plane wave front in 1D with amplitude 1 and initial reference phase $\phi = 0$ is represented by the following complex expression.

$$\psi_0(x) = \exp(2\pi ik.x) = \cos(2\pi k.x) + i\sin(2\pi k.x) = Re + Im \quad (1)$$

Where, $\psi_0$ is the plane wave function, $k = \frac{1}{\lambda}$ is the wave vector, and $x$ is the spatial dimension over which the amplitude of stationary wave is spread with cosinusoidal oscillation. The wave nature of light is observed e.g., during the passage of light through the slits under appropriate experimental conditions such as coherent illumination, monochromaticity, relative dimension of illumination wavelength, slit and periodicity. In case of electron, which is a mater wave, the wave nature manifests in the form of diffraction pattern of crystal, Fresnel edge pattern and off-axis electron holography fringes.

The Fraunhofer pattern in diffraction plane can be calculated within the celebrated Fresnel-Huygens construction. The origin of diffraction phenomena is due to the superposition of various spherical wavelets emanating from object forming envelope wave with modulation in amplitude and phase along different momentum direction. However, the geometry of interference and associated phase correlation between various wave vectors is different for the near-field and the far-field regimes (sec. S1.1 & 1.2). There exists analytical expressions for various types of aperture geometries through the evaluation of the Fraunhofer integral [21]. The Fraunhofer pattern can also be calculated by absolute Fourier transformation (abs-FT) of the real aperture function [Fig. 1], but the intensity pattern is given as a function of frequency $k$ of the plane wave basis and need to be calibrated either with the scattering angle $\theta$ derived from analytical formula or through a calibrated diffraction pattern.

The Fourier based method is analogous to the physical picture of Abbe's hypothesis of image formation, where the point of interaction between probe and specimen generates outgoing waves with continuous range of frequencies, waves with higher frequency propagate along higher scattering angles (sec. S1.3). Whereas, in Fourier method, the object function is expanded into various frequencies of plane waves with no information on the scattering angle. Both the scattering processes involve presence of various frequency components of the outgoing waves that is reminiscent to the picture of Bloch wave propagation of probe electrons in a periodic crystal but with discrete range of frequencies having symmetry with the crystal periodicity.[30] The inverse Fourier transformation returns back the aperture function, if the $\boldsymbol{Re}(x, k)$ and $\boldsymbol{Im}(x, k)$ components of the Fourier waves are known. The phase of the wave function that carries the object information, can be calculated by arctan function, i.e., $\tan^{-1}\left(\frac{Im(k\ or\ x)}{Re(k\ or\ x)}\right)$ corresponding to diffraction plane and image plane with variables $k$ and $x$, respectively.

For atomic systems, the electrostatic potential replaces the slit object function. The FT of isolated and periodic atomic potential in the diffraction plane is the atom scattering factor $f(k)$ and structure factor $F(g)$, respectively. The transmitted wave function in this case can be derived from the Schrödinger integral equation involving interference between incident plane wave and outgoing spherical wave with amplitude factor given by the atom scattering factor (supp. Sec. 1.4) [23,31]. The image intensity of atom can then be calculated following Huygens's principle in Fraunhofer approximation as given in Eq. S16 with the consideration of resolution limiting aberrations acting as a coherent envelope. The phase contrast image calculated varies weakly with atomic number and the peak phase shift $\varphi_{max}(rad)$ follows ~ $Z^{0.6} - Z^{0.7}$, where $Z$ is the atomic number [32]. This is possible as the formulation of scattering factor has the information on the atomic potential which corresponds to both crystallographic and object phase information. The strength of the scattering potential or object phase will only modify the magnitude of specific intensity pattern due to crystallographic phase. Generally, this so-called phase information is lost in the experimental diffraction pattern and only the absolute of ***Re*** and ***Im*** components or abs-FT are known. However, it is shown in sec. 4.B that it is possible to retrieve the image intensity function of certain types directly from the diffraction pattern. Once the image function is known, the object phase can be determined according to the standard procedures as described in sec. C or by the alternative method presented in sec. 4.A.

### B. Zernike phase object and weak phase object approximation (WPOA)

The method described here yield image intensity pattern at or near the Gaussian image plane. The definition of weak phase object is introduced here and its implication on the recorded intensity pattern. For pure phase or weakly scattering object the one-dimensional object function is mathematically represented by a complex function of the form $F(x) = e^{i\phi(x)}$ or $\sim 1 + i\phi(x)$ for small $\phi$, where $\phi$ is a real function corresponding to discrete or periodic transparent object and is known as weak phase object approximation (WPOA). The effect of Zernike phase plate modifies the intensity of the object wave that depends linearly on the phase change due to object according to $I(x) = 1 \pm 2\phi(x)$. This is the foundation of the Zernike's phase contrast theory.[21] In fact, the object function is real, it is the replica of the object carried by the probe plane wave that is a complex wave function (similar to the Fourier waves propagation with the information of the object function) and absolute of this function after interaction is imaged on the recording device. Therefore, in the recorded image pattern the information on the object function or crystallographic phase is preserved. The phase corresponding to OEW may be determined from this pattern using arctan function if the corresponding object wave is known.

However, similar approach for atomic system incorporated the change in phase in the plane wave illumination in terms of change in magnitude of the wave vector ($\Delta k = 1/\Delta\lambda$) [22,23]. This brings about transmission function and under axial imaging condition is further convoluted with the point spread function (PSF) in the image plane or multiplication with the phase contrast transfer function (PCTF) function which acts as a coherent envelope in the diffraction plane incorporating the effect of aberration phase shift to form the final image [19,22]. The intensity depends on the phase change due to object potential linearly after ignoring the higher order interaction terms (Eq. S21) that is similar to Zernike like transfer if the complete envelope functions is set to 1. Considering that the spatial resolution criteria of the instrument is met, again the information on crystallographic phase is not lost in the image intensity pattern. However, if the WPOA approximation is not considered, then the information related to object function $\phi(x)$ and complete phase information is lost [Sec. S1.4.2]. Kindly note that, the phase shift due to aberration cannot be added in the trigonometric function in diffraction plane as that shifts the wave, rather it is used as coherent enveloped or a frequency filter function. In real space, the *psf* due to PCTF gives weight to the intensity (e.g. Scherzer transfer will have maximum weight for optimum value of spherical aberration and defocus) that depends on the integration value of PCTF over the limits of reciprocal vector acting as a frequency filter and other coherent aberration e.g., aperture function determines the broadening or full width at half maximum (FWHM) of the potential function [19]. As the convolution procedure changes the magnitude of the resultant function significantly, a flux balance approach is helpful to observe the decrease in intensity and increase in FWHM due to aberration compared to ideal image free from any aberration (Fig. S7).

However, the above description of WPOA does not draw any analogy in terms of interference geometry between Gabor's in-line holography, Fresnel diffraction geometry and defocus HRTEM image except transmission function between in-line holography and transmission function derived based on change in magnitude of electron wavelength (compare Eq. S24 & S25). However, instead considering the change in momentum vector direction due to interaction with the object potential provides insights in all the three pictures in terms of geometry of interference (sec. S1.1 and S1.2).

### C. Various existing procedures on HRTEM focal series reconstruction

Reconstructions methods associated with the HRTEM through focal image series aim to retrieve the unknown phase $\phi$ of the OEW function of the form $\psi_i = A(x, y)e^{i\phi(x,y)}$. The OEW function can be used to interpret the object structure $f(x, y, z)$ based on the model methods as described in the previous sections. Generally, the illumination is considered to be monochromatic and for semi-monochromatic wave, partial coherence theory appears into the stage. While theoretical model behind off-axis electron holography technique to retrieve the OEW function is straightforward (sec. S2.2), but the methods corresponding to inline holography using through focal HRTEM image series are much more elaborate involving intensive data refinement and fitting procedures [17]. This is due to the nature of the fitting equation considered to retrieve the $\psi_i$ that attempts to reduce the effect of complex conjugate $\psi_i^*$ and non-linear image components $\psi_i\psi_i^*$ from the recorded image set [Eq. 5 & 7].

These exist few different focus variation methods e.g., Wiener formulation by Schiske, 3D paraboloid method (PM), maximum likelihood method (ML) and various numerical schemes associated with them [2,3,33]. Saxton showed the equivalence between the different reconstruction techniques in terms of equivalence in the form of restoring filter applied to the intensity expression in Fourier space to retrieve the wave function. The simplified form of the restoring filter is given as

$$r_n(k) = \frac{1}{N} \exp\{i\gamma(k)\} \tag{2}$$

and the associated aberration function

$$\gamma_n(k) = \pi C_s \lambda^3 k^4 - \pi \lambda z_n k^2 \tag{3}$$

finally, the restored wave function is written as

$$\psi(k) = \sum_n I_n(k) r_n(k) \tag{4}$$

Where, $C_s$ is the third order spherical aberration, $\lambda$ is the wavelength, $z_n$ is the defocus corresponding to the $n^{\text{th}}$ image, and $k$ is the spatial frequency. The application of restoring filter is similar to deconvolving the effect of aberration. The inverse-FT of Eq. 4 will return the wave function in the image plane. This is only possible if one starts with the intensity pattern recorded in the image plane with sufficient spatial resolution and during inverse-FT, spatial information on the crystallographic phase is preserved through $Re(x, k)$ and $Im(x, k)$ of Fourier waves. The pattern of wave function in terms of phase $\phi$ and amplitude $A$ thus obtained required to be fitted with the model calculation for further interpretation.

However, Saxton derived the final form of the wave function without considering any restoration filter from a different start in the form of the intensity expression as given in Eq. 5. The approach provides insight into the effect of restoration on dispersing the effect of conjugate wave function and non-linear image component. Under perfect coherence, the image intensity at some focus *z* near the Gaussian image plane is written as

$$i(x,z) = |1+\psi_i|^2 = 1+\psi_i(x,z)+\psi_i^*(x,z)+h(x,z) \quad (5)$$

Where, $\psi_i(x,z)$ is the wanted part or the OEW, $h(x,z) = \psi_i(x,z)\psi_i^*(x,z)$ and $\psi_i^*(x,z)$ are the unwanted non-linear term and complex conjugate, respectively. $\psi_i^*(x,z)$ is the twin image of $\psi_i(x,z)$. Kindly note that the reference wave $\psi_0$ is set to 1 by considering axial illumination ($k_0 = 1$) in the above expression.

The 2D Fourier transformation of Eq. 5 and then adding explicitly the dependence on defocus results in the following expression.

$$I(k,z) = \delta(k)+\psi(k)\exp(\pi i\lambda zk^2)+\psi^*(-k)\exp(-\pi i\lambda zk^2)+H(k,z) \quad (6)$$

Where, $H(k,z)$ is the non-linear component that describes the autocorrelation between the two linear terms in the reciprocal space. The paraboloid method as proposed by van Dyck can be derived from the above expression by taking 3D-FT with respect to defocus where the wave function and its complex conjugate follow the reflected parabola from the reference diffraction plane [34].

Now, within the coherent detection i.e., multiplying by corresponding phase conjugate and summing over N images after assuming $k$ is non-zero which allows to omit the delta function and for a constant $k$, the image intensity becomes

$$\sum_n I_n \exp(-\pi i\lambda z_n k^2) = \psi\sum_n 1+\psi^*\sum_n \exp(-2\pi i\lambda z_n k^2)+\sum_n H_n \exp(-\pi i\lambda z_n k^2) \quad (7)$$

From the above expression, it was concluded that the wave functions $\psi$ and $\psi^*$ accumulate to N and $\sqrt{N}$ times to its original value, respectively. The non-linear term $H_n$ behave randomly with the defocus. Thus, by dividing the above equation by N, the wanted OEW function may be recovered. The Eq. 7 is widely considered in almost all the through focus image series OEW reconstruction. And all the efforts on developing image reconstruction codes primarily deals with eliminating the non-linear and complex conjugate terms and finding a best fit with the model calculation. However, the presence of complex conjugate and non-liner terms will always be present depending on the extent their weights are subdued.

Nonetheless, there exists a contention, after complete evaluation of Eq. 5, yields the following final form

$$I_{in\ line} = |\psi_0 + \psi_i|^2 = A_0^2 + A_i^2 + 2A_0A_i cos(\phi_i - \phi_0) \quad (8)$$

Where, $\psi$ is replaced with the wave function of the form $A(x,y)e^{i\phi(x,y)}$, describing the image intensity pattern based on self-interference between reference incident and scattered waves within the picture of single electron wave interference phenomena [Eq. S49 & S50]. The expression has the similarity with the wave interference between the reference and object waves in off-axis electron holography (sec. S2.2). However, in off-axis geometry, there is an additional phase term $Qx$ due to wave interference at an angle that gives rise to spatial modulation in the interference field [35]. Moreover, Eq. 5 is in an intermediate state, which is used to eliminate the effect of twin image and non-linear terms by working in the diffraction plane [sec. S2.1]. However, the existence of the final form of the expression implies that by fitting the intensity equation alone and evaluating the phase term should in principle allow to extract the relative phase change from the image plane. For more details on associated twin image wave functions and applicability of Eq. 8, see sec. 4. The results based on above schemes can be found in Ref. [6,17] and does not show any systematic trends with the sample thickness.

Finally, the approach based on partial coherence theory considers the effect of finite source size, chromatic defocus spread, current voltage fluctuation of the instrument, objective aperture size and wave aberration function [1,36,37]. The reconstruction method based on partial coherence theory is known as iterative linear restoration which addresses the residual non-linear term. By repeated application of linear restoring filter from the subtraction of the calculated non-linear term improves the initially guessed wave function. Various derivations available based on partial coherence theory is given in sec. S2.4. One can notice that the coherence and interference phase shift in the formalism has the origin in convolution procedure in real space and cross correlation in Fourier space.

## 4. Alternative methods on the phase retrieval

In this section the alternative proposals are introduced for the retrieval of the $\phi$ and $A$ of OEW function from the image and diffraction planes, respectively. The methods are straightforward and do not require through focal image series acquisition similar to off-axis electron holography where only single image embedded with the hologram is sufficient. Addressing the effect of large defocus together with various coherent aberration envelopes, and thickness on the rich variations in phase and image intensity pattern by the alternative procedure will be part of a future discussion as this requires a different viewpoint from the traditional approach due to unique experimental observation overlooked in the past.

### A. Recovering phase from the HRTEM image intensity pattern

The method described here works on atomic resolution HRTEM image recorded under suitable imaging condition i.e., with a particular combination of spherical aberration coefficient $C_s$ and defocus $\Delta f$ that sets the optimum contrast and resolution.

The object phase can be recovered by applying Eq. 9 as given below which is a modified form of Eq. 8 describing the image intensity pattern in HRTEM within few nanometers from the Gaussian image plane.

$$I_{in\ line}(x,y) = |\psi_0 + \psi_i|^2 = A_0^2 + A_i^2 + 2A_0A_i\ sin\{\phi_i(x,y)\}$$

$$= \alpha I_0 + \beta I_0 + 2\sqrt{\alpha I_0 \times \beta I_0}\ sin\{\phi_i(x,y)\} \quad (9)$$

Where, $I_0$ is the mean vacuum intensity. The factors $\alpha$ and $\beta$ represent the fraction of direct and scattered part of the intensity and can be determined by analyzing the image pattern, where $\alpha + \beta = 1$. Typical values of $\alpha$ and $\beta$ are found to be ~ 0.88/0.95 and 0.12/0.05, respectively from the experimental images of $MoS_2$/BN.

And the intensity expression corresponding to off-axis electron hologram is given by

$$I_{off\ axis} = 1 + a^2(x,y) + 2a(x,y)\cos\{2\pi Qx + \phi(x,y)\} \quad (10)$$

The appearance of sinusoidal function in Eq. 9 in contrast to Eq. 8 is due to OEW phase term has a relative phase term with respect to vacuum phase $\phi_0$ in addition to $\phi_i$ due to object potential in case of HRTEM. This phase difference is absent in off-axis electron holography [compare Eq. 9 & 10] and is generally considered to be $\pi/2$ i.e., $\Delta\phi = \frac{\pi}{2} - \phi_i$, that sets the vacuum phase value to zero. In case of off-axis geometry, the two halves of the wave on either side of the bi-prism carries the same vacuum reference phase term and gets eliminated in the final intensity except part of the wave that carries the object phase information [sec. S2.2]. For an intuitive physical picture on the origin of such a $\pi/2$ phase shift between diffracted and primary incident waves based on interference geometry is provided in sec. S 2.2.1 [Fig. S11]. The modifications incorporated in Eq. 9 are essential and yield excellent results from various experimental images with different types of atoms present for a given crystal. For more discussion on the factors $\alpha$ and $\beta$ and particular use of amplitude terms and various forms of Eq. 10 and its relationship with experimental image see sec S5.

Now, both the equations are founded on the physical picture of self-interference of single electron wave function involving interaction in the form of phase change with the object potential [20]. The interaction is mostly elastic due to fast probe electron and small probability of inelastic interaction is useful for the analytical techniques. The primary difference between Eq. 8 & 10 is that additional phase term of $2\pi Qx$ that appears in the trigonometric function in off-axis electron holography due to wave interference at an angle (sec. S2.2). This particular phase term oscillates with the spatial coordinate $x$ over the field of the electron hologram. The phase term $\phi(x, y)$ corresponding to OEW phase is acquired by another half of the wave while passing through the sample. This OEW phase manifests as holographic fringe bending in the image plane, e.g., relative fringe bending between vacuum and MgO crystal and differently striped thickness of object structure, that allows to determine the mean inner potential (MIP) at medium resolution [4,38].

Another difference between the two holography techniques is the twin image. It is well known that for in-line holography separation of twin image is an issue whereas for off-axis geometry they get separated at $\pm Q$ in the frequency space. However, in the image plane they superimpose on top of each other for both the techniques. In standard practice of OEW reconstruction, the Eq. 9 & 10 are generally written in an intermediate state (see Eq. 5 for in-line and Eq. S54 for off-axis) having both OEW function $\psi$ and its conjugate $\psi^*$ in consistent with the holographic principle of image formation [21]. The motivation behind this was based on the concept that the twin image components can be separated from the direct component (DC) if allowed them to propagate along the scattering direction i.e., from the image plane to the diffraction plane and emphasizing on the direct retrieval of the wave function. However, it is argued here that this does not poses any issue if one wishes to work in the image plane of the intensity pattern to retrieve directly both the phase and amplitude of OEW. The choice and limitations between the diffraction plane and image plane can be understood by Heisenberg's uncertainty principle, i.e., only momentum and position variables of the OEW are known in diffraction plane and image plane at a time, respectively.

Now, from the Eq. 9 & 10, it is evident that the phase information is preserved in the final form of the intensity equation that is responsible for the intensity modulation as described by the trigonometric function. At the same time, no information on the wave functions or twin images and DC component is available. The application of final form of intensity expression instead of wave function based intermediate state of equation can be guided by Born rule in Quantum Mechanics. Born rule states that it is the probability density $\psi\psi^*$given by the square of the probability amplitude of particle's wave function $\psi$ that is real and observable quantity during the measurement and not the associated complex wave function which is functions as a state vector [39,40]. There exists controversies in the literature on the measurement of such state vector which is a complex quantity but can be constructed based on the information of probability density [41–43]. Born rule is an important link between the abstract mathematical formalism of quantum theory based on the complex wave function and the experimental measurement and constitutes an essential part of the Copenhagen interpretation of quantum mechanics [44]. Therefore, the final form of the intensity equation is equivalent to measuring the total probability density due to contribution from all the three components of wave functions, i.e., DC part, wanted wave function $\psi$ and its complex conjugate $\psi^*$. And in the process of recording, the information on the wave functions is lost, recovering of which is emphasized in all the reconstruction approaches through intermediate state of the equations. Retaining the phase term is inherent to self-interference and resulting modulation in intensity pattern and similar to the off-axis electron holography fringe bending due to object potential in the image plane. The predicament of evaluating the local phase information from real space image in terms of fringe bending primarily comes from the restriction on spatial resolution and will be addressed in the forthcoming discussion along with additional aspects already mentioned earlier. In standard practice, the image is Fourier transformed and then a digital aperture function is used in that plane that takes care of the resolution. The use of intermediate state of the equation to disperse the twin images and DC part in the diffraction plane made the earlier methods elaborate and complex for HRTEM reconstruction where twin images overlap on the same reciprocal space and required to enhance the weight of $\psi$ over the $\psi^*$ and $\psi\psi^*$ terms through summation over many images. Though the information on the wave functions is lost, however, the phase term is preserved in the final state of the equation which are the same for both $\psi$ and $\psi^*$ but having opposite sign.

Now, the twin images are exact copy of each other and superimposes at the Gaussian image plane and do not cause any loss in information in terms of phase and amplitude. This can be understood with the help of off-axis electron holography, where inverse FT of each side band preserve the same intensity pattern. In inline holography, the twin images propagate along the opposite direction with defocus similar to the parabola picture of OEW reconstruction where complex conjugate pairs lie on a reflected parabola with respect to the diffraction plane and focus variation. This picture is reminiscent to the transactional interpretation of quantum mechanics where the wave function and its complex conjugate experience phase change in the opposite directions equivalent to the forward and backward propagation in time [45,46].

Now proceeding for the experimental reconstruction of OEW function, at first the phase can be determined directly by applying Eq. 9. If the quantity inside the arcsine function is more than 1, then this should be divided by 1 and the computer program should read the quotient as phase jump by the amount (quotient × $\pi/2$) plus the phase corresponding to the remainder and then calculate the total phase change and corresponding counts of the atoms accordingly. This problem does not appear for one-layer thick $MoS_2$ and BN but encountered in 3D crystal like ZnO where thickness includes several atoms in the column. Next, the amplitude corresponding to $\psi$ needs to be evaluated from the image intensity. The total amplitude of the recorded image is simply the square root of it but not equal to the amplitude of the single OEW function. As already mentioned, that the total amplitude in image intensity pattern has three components, direct component (DC), and two twin image components for off-axis and DC and only diffracted components in case of in-line holography. In case of in-line holography the twin image components overlap with the DC part. The DC component has approximately one order of magnitude higher total intensity compared to individual twin images from SBs and diffracted FT spots as measured directly from the FT of the images by placing a region of interest around various spots and SBs and evaluating the sum. Thus, the intensity of OEW function is approximately given by the factor $\beta$ for HRTEM and this will be further divided in halves between two twin image components in case of off-axis holography. The amplitude of $\psi_{HRTEM} = \psi_0 + \psi_i$ is given by the square root of $I_{in\ line}(x, y)$ and only amplitude component $A_i$ associated with trigonometric function is shown in reconstructed wavefunctions, where $A_i$ and $A_0$ are the magnitudes of individual diffracted and reference complex wave functions, respectively. Details on the evaluation on the amplitude from the intensity pattern, discussion on the interpretation of amplitude of isolated wavefunction and on the appropriate form of conventional intensity expression of atomic resolution off-axis hologram and HRTEM are given in sec. S2.2. Kindly note that the subtraction procedure to adjust the intensity value is automatically taken care by Eq. 9 while calculating the phase. Therefore, it is possible to reconstruct the OEW function from the information obtained from the measurement as described above i.e., phase and amplitude.

We have analyzed selected experimental images of $MoS_2$, BN and ZnO using Eq. 9 and the reconstructed phase and amplitude images are presented in Fig. 2 & 3 for $MoS_2$ and BN, respectively. One can see from the phase images that the vacuum phase is zero and phase values are high and low depending on the brightness in the dots which represents the periodic arrangement of atoms in the lattice. Theoretical peak phase shift values extracted from Ref. [32] for the resolution range 1 - 0.5 Å is given in Table S1 for the atoms of interest in the present investigation. According to this table, the peak phase shift $\varphi_{max}(rad)$ follows ~ $Z^{0.6} - Z^{0.7}$, where $Z$ is the atomic number [32]. The exponent of Z increases with increasing spatial resolution results in higher peak phase shift value for a given atomic number. Though the trend can be complicated depending on the valence electron filling and for specific atoms with higher $Z$ can have smaller contrast compared to atoms with lower Z next to each other in the periodic table. From the present reconstruction method, a peak phase shift value of ~ 0.56 and 0.45 rad are obtained from the Mo and S atom positions, respectively throughout the image, which are little higher than the theoretical values. Kindly note that for $MoS_2$ there are two S atoms along the [0001] projection, therefore, the phase shift values are almost double compared to single S theoretical phase shift value. The values are 0.06 and 0.21 rad, for B and N atoms, respectively which are close match with the theoretical estimation. Result on ZnO and bi-layer $MoS_2$ are given in Fig. S13 & S14. For ZnO thin film, as the sample thickness increases from the vacuum edge of the specimen, systematic increase in phase and atoms numbers can be observed.

The method described above using Eq. 9, in fact translates the intensity information into a phase value and an excellent trend is obtained for images with low and high Z compounds giving lower and higher phase shift values, respectively without any additional data refinement procedures. Reference vacuum is not required as one can approximate the mean of atomic resolution image intensity as mean vacuum intensity. Intensity pattern if at all present other than OEW, e.g., Fresnel fringe, electron interference fringe in off-axis electron holography should be removed from the image data before applying Eq. 9. These unwanted patterns are generally removed from the image before the final analysis in case of off-axis electron holography.

### B. Recovering image function from diffracted intensity by Fourier series expansion

In this sub-section the recovery of image function containing information on both crystallographic and object phase is described if the information is available only in the diffraction plane. The real image function $f(x)$ of both symmetric non-periodic and periodic in particular forms e.g., Gaussian, reciprocal etc. can be retrieved by cosine Fourier series expansion of the absolute FFT of such functions followed by summation over all frequencies. As already mentioned, that the absolute FFT and Fraunhofer pattern are equivalent to each other. The procedure is similar to the zero phase reconstruction using Patterson function in X-ray crystallography applicable to smaller size molecules [24,25]. Result is presented for the example Gaussian function in 1D both in isolated and periodic form (Fig. 4). Illustrations of other type of functions and 2D forms are given in supp. sec. 6.

The unique feature underlying the workings of this reconstruction method is that for most functions the integration of imaginary part over spatial parameter or abs-Im component is negligible compared to real part or abs-Re [Fig. S9]. Therefore, the original function $f(x)$ can be retrieved completely by the following real cosine Fourier series alone. If $F(k)$ is the absolute FFT of $f(x)$, then $f(x)$ can be retrieved from the $F(k)$ directly via following real Fourier series expansion

$$f(x) = absolute\left(\sum_{n=-k}^{n=k} \frac{1}{n} C_k \cos(2\pi k x)\right) \quad (11)$$

Where, $n$ = number of data points in frequency axis, $k$ is the frequency, $C_k$= absolute of Fourier transformations or diffracted intensity at some frequency $k$, and $x$ is the 1D spatial coordinate over which real image function will be defined.

For periodic function in 1D, $f(x) = f(x \pm px)$, where $p$ =0, 1, 2, …, one need to ensure that the range of $n$ should be $= \pm \frac{2p}{M}$, where $M$ = number of data points both in real and diffraction space. In case of 2D isolated function the Eq. 11 will have another summation over $n_y$ for the second orthogonal axis and the modified Fourier series equation for isolated and periodic 2D function are given in supp. Sec. 7 [Eq. S85 & S86], where the functional form of cosine functions changes depending on the isolated 2D and periodic 2D functions in order get an exact fit.

The method described above works well for both isolated and periodic functions. From the image intensity function one can then apply the first method to retrieve the OEW and associated object phase related to potential. However, if there is any inhomogeneity in the distribution of periodic function e.g., dopants, edges, interfaces etc. then the accompanying information will be required from the image intensity pattern and the method described in previous section works suitably.

## 5. Conclusions

In conclusion, alternative reconstruction methods to retrieve the phase and amplitude of OEW in HRTEM imaging are introduced. The first method is based on directly applying a modified HRTEM intensity equation to retrieve the phase of OEW from the image. The reconstruction results in terms of peak phase shift values are in excellent agreement with the theoretical estimation. The second method described is applicable for retrieving the image intensity function from the information available in the diffraction plane for both isolated and periodic functions.

## Acknowledgement

The authors at JNCASR sincerely acknowledge ICMS, JNCASR for the financial support. Prof. Ranjan Datta would like to give a special thanks to Deutsche Forschungsgemeinschaft (DFG) for providing funding for an illuminating and exceptional research visit in TU-Berlin in the year 2016.

**Figures:**

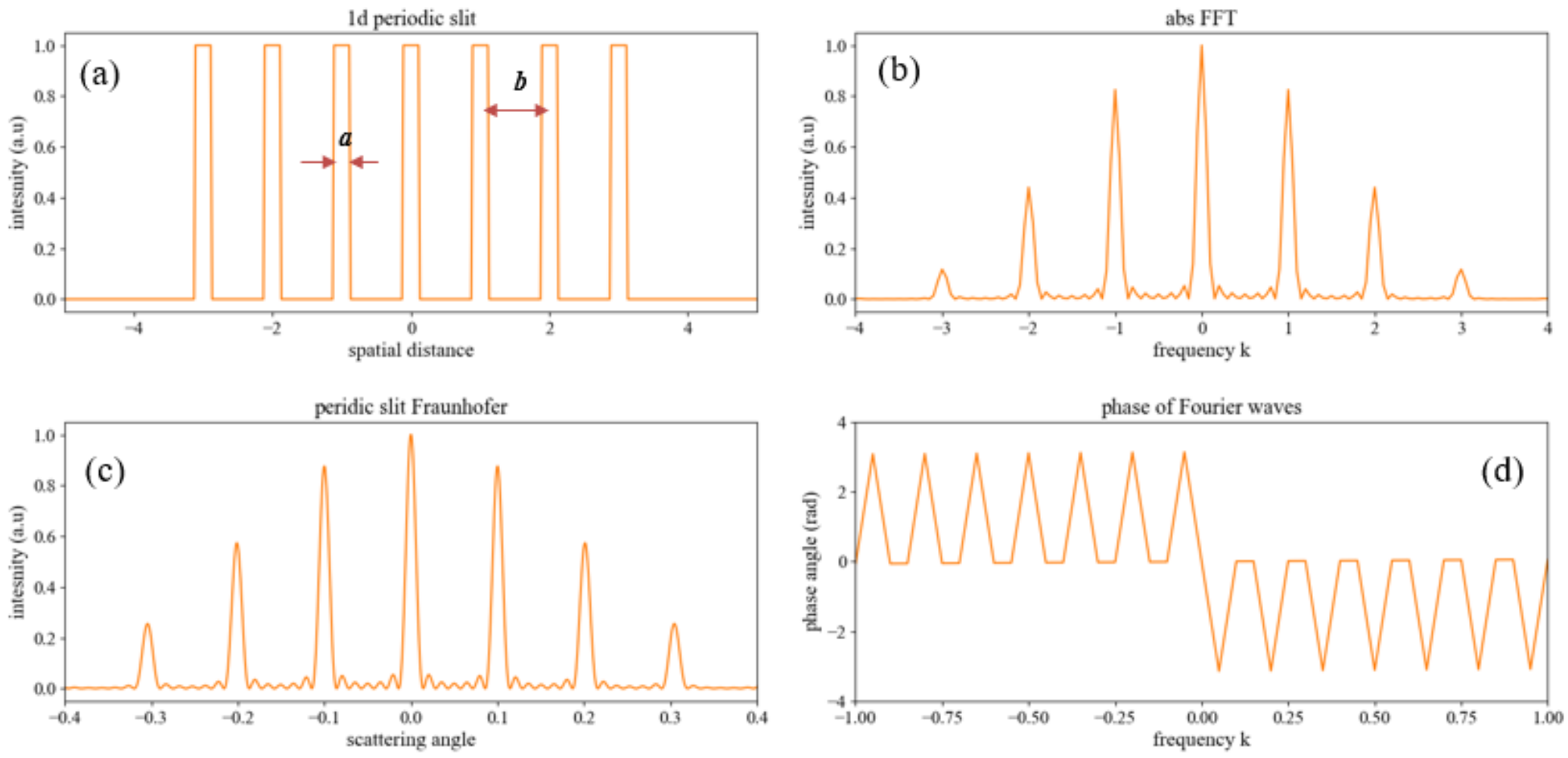

***Fig. 1.*** *(a) Example seven slit periodic object with slit width **a** and slit periodicity **b**. Fraunhofer pattern calculated (b) based on Fourier transformation and (c) by applying analytical formula which is based on physical picture of constructive and destructive interference between waves along various direction. The phase angle for each Fourier frequency is shown in the graph (d).*

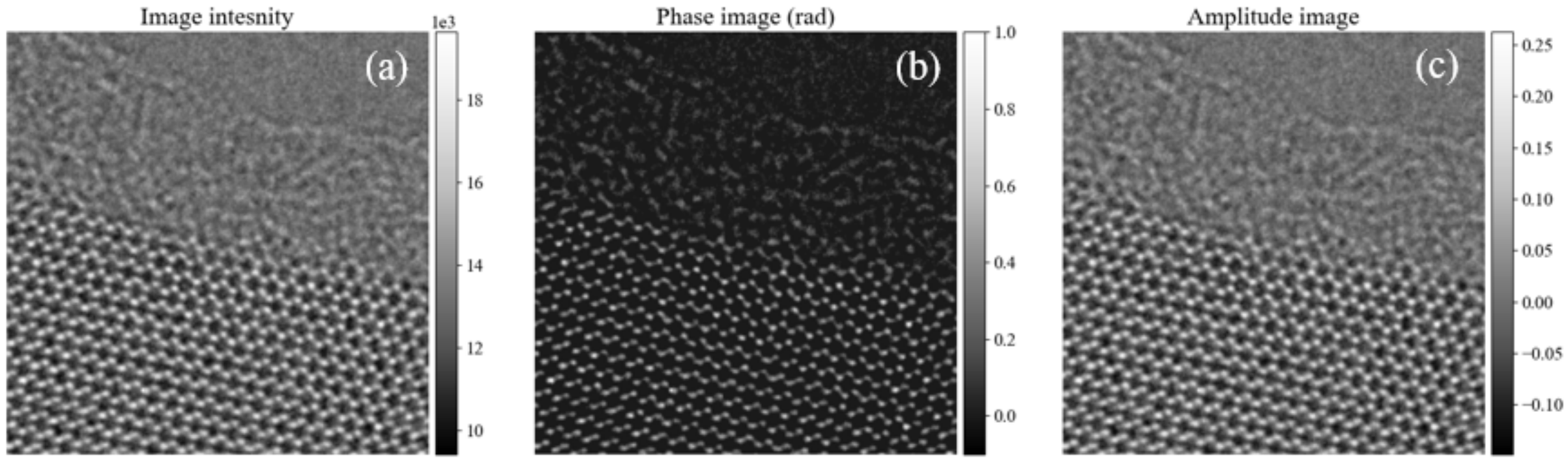

***Fig. 2***. *(a) HRTEM image of MoS2 recorded under negative* $C_S = -35\ \mu m$ *and positive defocus of* $\Delta f = 8$ *nm. (b) phase and (c) amplitude of the OEW. A peak phase value of ~ 0.56 and 0.45 rad are obtained for Mo and two S atoms, respectively.*

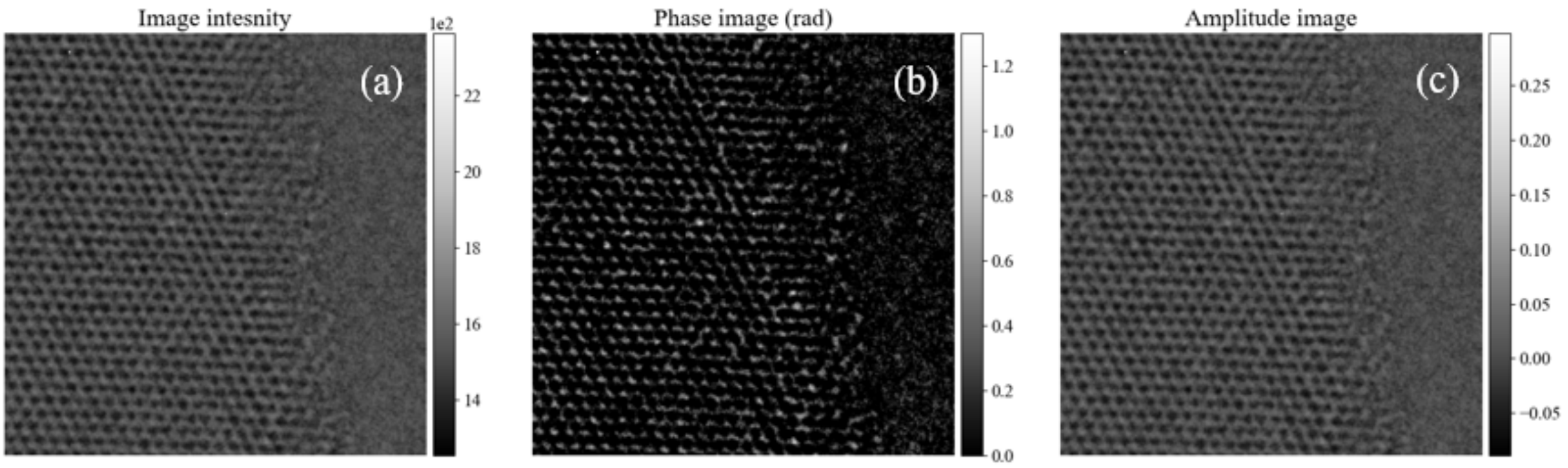

***Fig. 3***. *(a) HRTEM image of BN monolayer recorded under negative* $C_S = -35\ \mu m$ *and positive defocus of* $\Delta f = 8\ nm$*. (b) phase and (c) amplitude of the OEW. A peak phase value of ~ 0.06 and 0.21 rad are obtained for B and N atoms, respectively.*

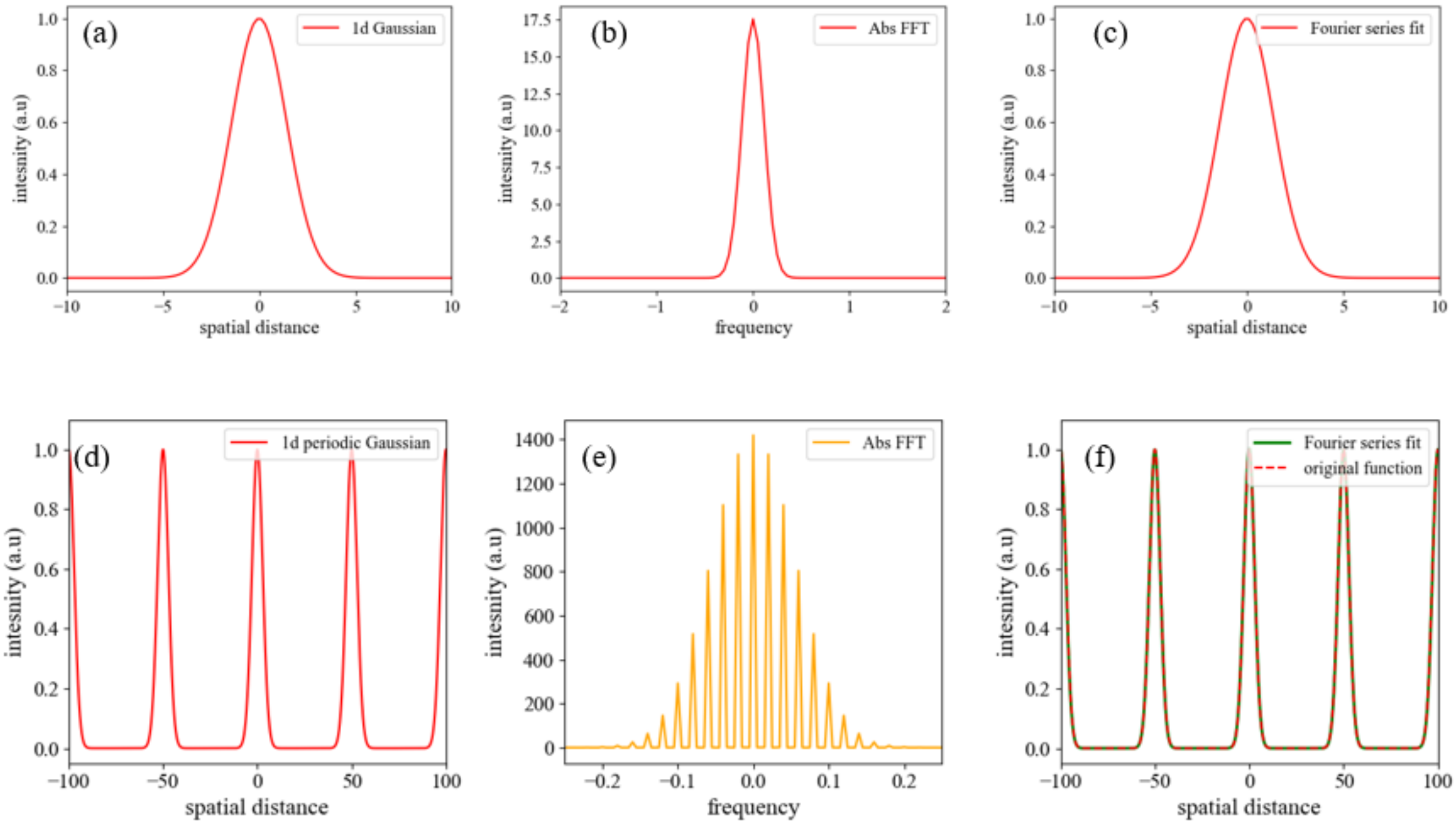

***Fig. 4***. *(a) single Gaussian function in 2D, (b) absolute FFT of the function and (c) Fourier series expansion followed by summation over all the frequencies. The result based on 1D form of the function is displayed in the insets. (d) Periodic Gaussian function in 1D, (e) absolute FFT of the function and (f) Fourier series expansion*

*followed by summation over all the frequencies. The result based on 2D form of the periodic function is given in supplementary document.*

## Supplementary Document

# Insights and alternative proposals on the phase retrieval in high resolution transmission electron microscopy

Usha Bhat, and Ranjan Datta[1,2,*]

[1,2]*International Center for Materials Science, Chemistry and Physics of Materials Unit, JNCASR, Bangalore 560064.*

Reconstructions methods based on experimental high-resolution transmission electron microscopy (HRTEM) images aimed to retrieve the unknown phase ($\Delta\phi$) of the object exit wave (OEW) function of the form $A(x,y)e^{i\phi(x,y)}$ from the recorded images. The phase change in the probe plane electron wave is brought about due to interaction with the specimen electrostatic potential. The missing phase information if retrieved, can be used to interpret the object structure in terms of potential distribution. For semi-monochromatic wave, partial coherence theory is adopted [sec. 2.4]. There are couple of experimental techniques namely HRTEM and off-axis electron holography available to perform the essential experimentations both at medium and atomic resolution. While theoretical model behind off-axis holography technique to retrieve the phase (or complete wave function) information is straightforward [sec. 2.2], however, the methods corresponding to inline holography or HRTEM from through focal image series are not so and involved extensive data refinement and fitting procedures.[1] Therefore, at first a detailed description is provided the on various existing methods related to flow of phase information to the image and diffraction planes, existing reconstruction techniques in the context of HRTEM and off-axis electron holography and then add supplementary details on the retrieval of the phase and OEW directly based on alternative methods from the information recorded in image and diffraction planes, respectively.

## 1. Formation of image intensity patterns of objects at various optical planes due to change in phase in the probe illumination

### 1.1. Fresnel diffraction pattern

The Fresnel diffraction is a near field pattern perpendicular to an interface which is due to discontinuity in the scattering potential near an edge or interface.[2,3]

Huygens-Fresnel principle which is based on following two postulates

(i) Every point of a wavefront is a source of secondary disturbance giving rise to spherical wavelets and the propagation of the wavefront is regarded as the envelope of these wavelets.

(ii) The secondary wavelets mutually interfere.

Now applying the above principle which is based on the diffraction geometry as shown in Fig. S1, each surface element d$S$ of the incident wavefront $\psi_{in}$ generates a spherical wavelet contribute an amplitude d$\psi_{sc}(P)$ at a point P on the optic axis beyond the wavefront

$$\mathrm{d}\psi_{sc}(P) = -iA(2\theta)\psi_{in}\frac{e^{ikR}}{R}\mathrm{d}S \tag{12}$$

after integration over all surface Eq. 1 becomes

$$\psi_{sc}(P) = -i\int_{wavefront} A(2\theta)\psi_{in}\frac{e^{ikR}}{R}\mathrm{d}S \tag{13}$$

After choosing appropriate limit over radial extent, Eq. 2 becomes

$$\psi_{sc}(P) = i\frac{\lambda\psi_{in}^{0}}{r_0+R_0}e^{ik(R_0+r)} \tag{14}$$

The above Eq. 3 describes the propagation of spherical wave. In HRTEM image simulation similar spherical wave is used in superposition with the incident plane wave front in the formulation of transmission function based on a solution of Schrödinger equation in integral form.[2,4]

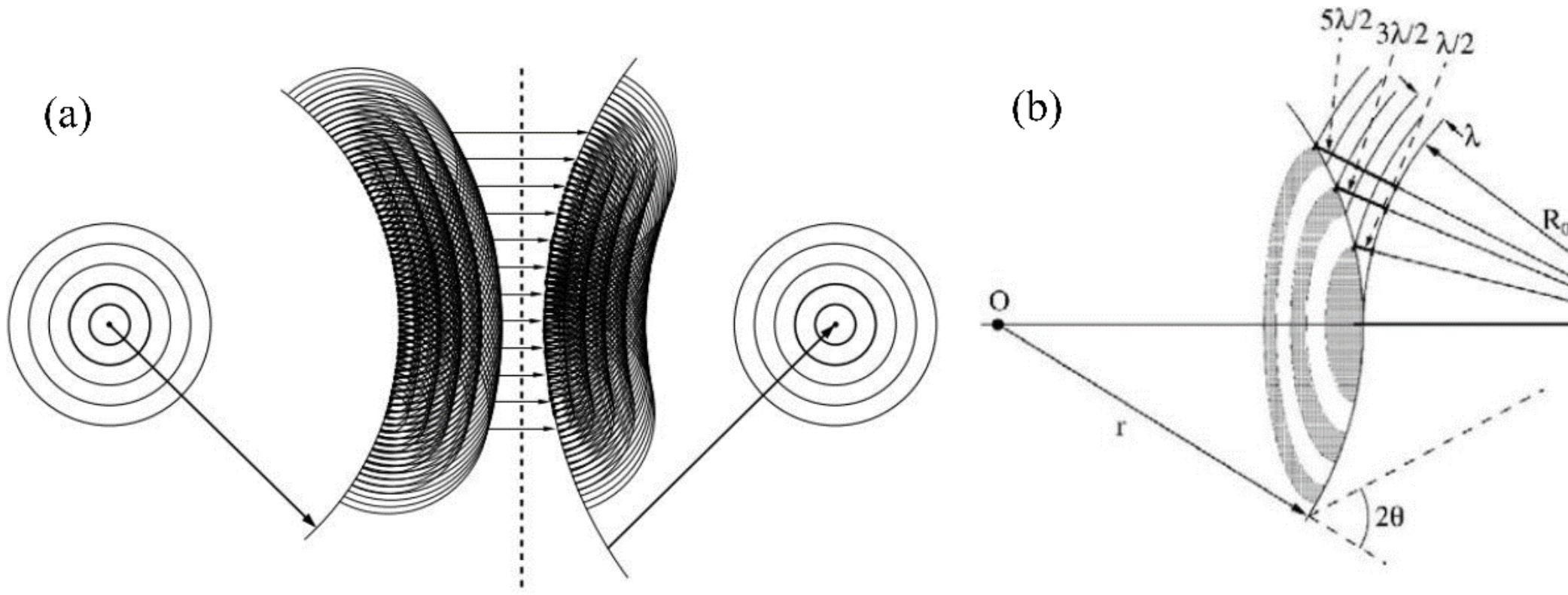

***Fig.S1.*** *(a) Geometry of Huygens principle for a diverging (left) and converging (right) wavefront. The action of lens is along the dashed line. (b) Construction of Fresnel zones considers the self-interference of the spherical wave while converging to a point P.*

The Fresnel diffraction pattern for an edge can be calculated using the following equation.

$$\psi_{SC}(P) = \frac{i\psi_{in}^{0}e^{ik(r_0+R_0)}}{2(r_0+R_0)}[C(X)+iS(X)]_{X_0}^{\infty}[C(Y)+iS(Y)]_{-\infty}^{\infty} \tag{15}$$

Where *C(X)* and *S(X)* are the Fresnel cosine and sine integrals and the plot of $C(X) + iS(X)$ is called a 'Cornu spiral'.

❖ Therefore, one can notice that the phase change in the propagating spherical wave is due to range of angular momentum vector directions (same path length lie on the surface of the sphere with respect to emitting point O) associated with a spherical wave (for plane wave it is only one momentum direction) and the self-interference while converging between those various angular momentum vector directions. Various angular momentum vector directions will acquire path difference between themselves due to *outside curvature of spherical surface* with respect to converging point P and the problem of interference is solved by well-known Fresnel zone construction. More precisely, the wave fronts geometries are spherical, parabolic, and plane surface for Rayleigh-Sommerfeld, Fresnel and Fraunhofer regimes, respectively. The rate of change of phase between wavevectors that is governed by the outward curvature of the wave front, is in different order between three different regimes. A similar picture is also captured in self-interference between probe and scattered waved that describes the HRTEM image intensity pattern (sec. 3.C, 4 and Eq. 5, 8, & 9).[5] The momentum vector for $\psi_0$ and $\psi_i$ are along the $k_0$ and $k$ directions, respectively. Kindly note that the interference geometry is different for Fraunhofer pattern and off-axis electron holography in comparison to Fresnel regime (sec. S1.2 & S2.2).

### 1.2. Fraunhofer diffraction pattern

Fraunhofer integral can be employed to calculate intensity pattern at far field for various geometry of slits. The integral yields an analytical expression which calculates the intensity of the pattern as a function of scattering angle.

The Fraunhofer integral for a rectangular aperture of sides 2a and 2b with origin at the center O of the rectangle and with $O\xi$ and $O\eta$ axis parallel to the sides is given as follows

$$U(P) = C \int_{-a}^{a} \int_{-b}^{b} e^{-ik(p\xi+q\eta)} d\xi d\eta = C \int_{-a}^{a} e^{-ikp\xi} d\xi \int_{-b}^{b} e^{-ikq\eta}\, d\eta \qquad (16)$$

And the intensity expression is given by

$$I(P) = |U(P)|^2 = \left(\frac{\sin kpa}{kpa}\right)^2 \left(\frac{\sin kpb}{kpb}\right)^2 I_0 \qquad (17)$$

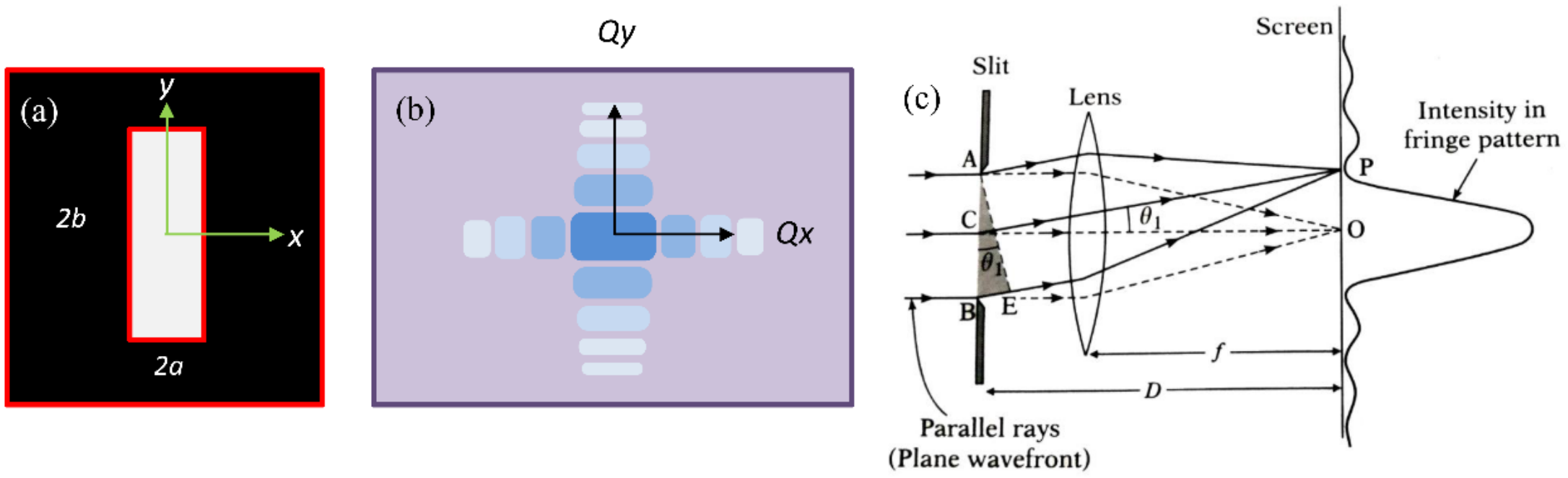

***Fig.S2.*** *(a) Geometry of rectangular aperture and (b) corresponding Fraunhofer intensity pattern. (c) Fraunhofer diffraction geometry.*

Experimental observation of example Fresnel pattern and far field Fraunhofer patterns are given in Fig S3.

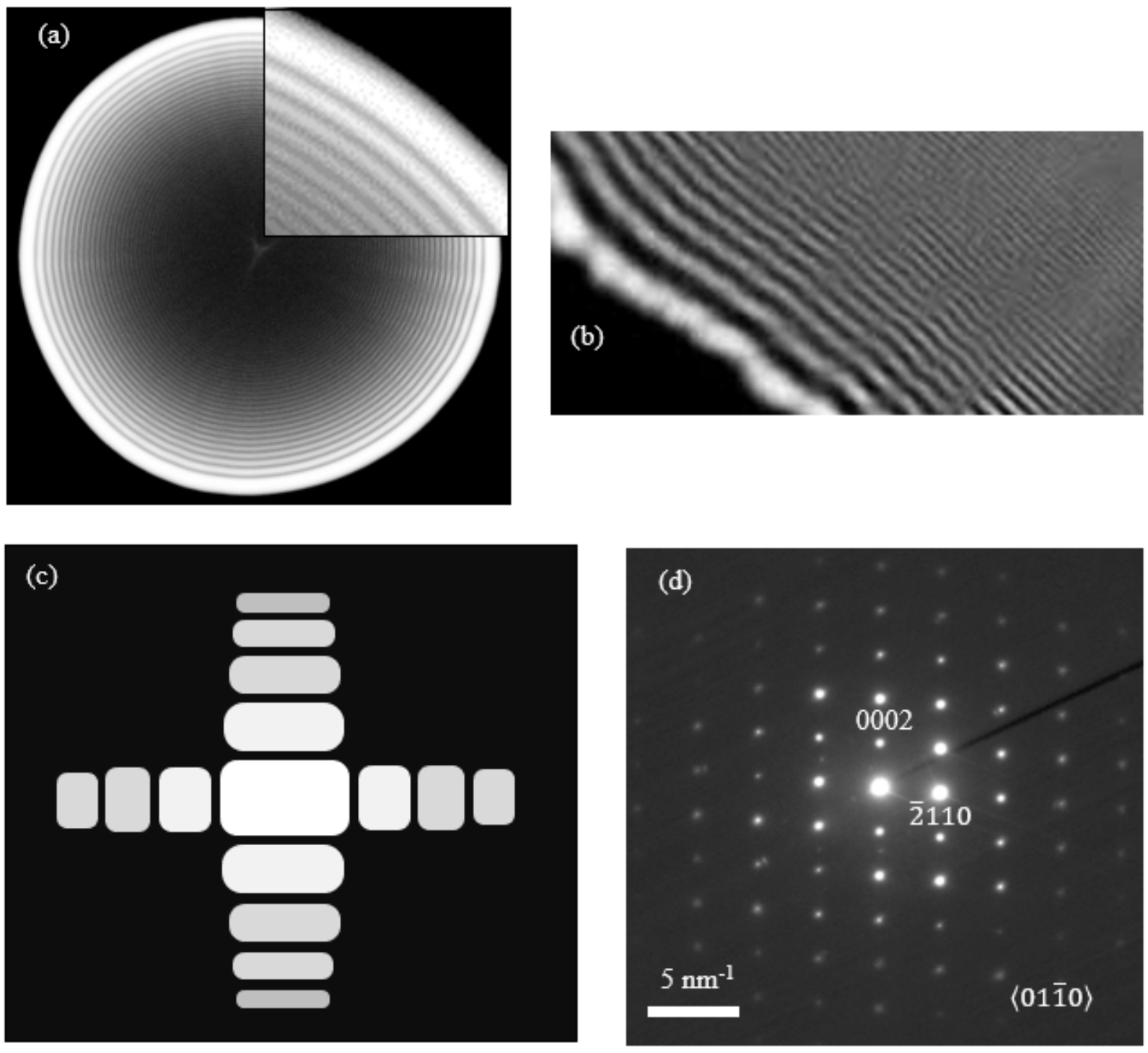

***Fig. S3.*** *Fresnel diffraction pattern as observed under slight defocus (~ 10s of nm) condition from the (a) edge of an aperture, and (b) thin specimen edge. (c) Far field Fraunhofer pattern from single slit and (d) electron diffraction pattern of ZnO Crystal along <01-10> Z.A.*

❖ As already mentioned, that the interference geometry for the far field pattern is different than the Fresnel zone construction. In this case the correlation of path difference between emerging waves from various spatial points at the aperture plane is considered. The path difference is not due to different momentum vector directions in this case rather spatial separation between two spatial points having same momentum or scattering directions. In case of off-axis electron holography the interfering waves have momentum direction mirror symmetry to each other (sec. S 2.2 and Eq. S 51).

## 1.3. Phase information propagation according to Fourier formalism or Abbe's hypothesis

In the preceding two sections, the propagation of phase information from the object plane to the near field regime (Fresnel) and far field diffraction plane (Fraunhofer) are briefly described. The change in momentum vector directions of the scattered/emergent waves and concomitant interference phenomena are at the center of calculating the intensity pattern. Here, object means slits or semitransparent object through which illumination wave with appropriate frequency can show off diffraction phenomena.

Now, method based on Fourier transformation (FT) which has close analogy with the Abbe's picture of image formation is described. The method is based on expanding the object function into plane wave basis with continuous range of frequencies. Though it returns the similar pattern as can be obtained through Fraunhofer integral but independent variables they represent are different. The absolute FFT is plotted against frequency of plane waves $k$ or reciprocal vector $g$, whereas the intensity pattern based on Fraunhofer integral is plotted against scattering angle $\theta$. Abbe's hypothesis is the connection between the two results in the sense that each Fourier wave with higher frequency will scatter at higher angle and is the basis of the famous resolution criteria which is limited by the aperture opening. Calibration is thus required to extract the useful quantities based on FT method.

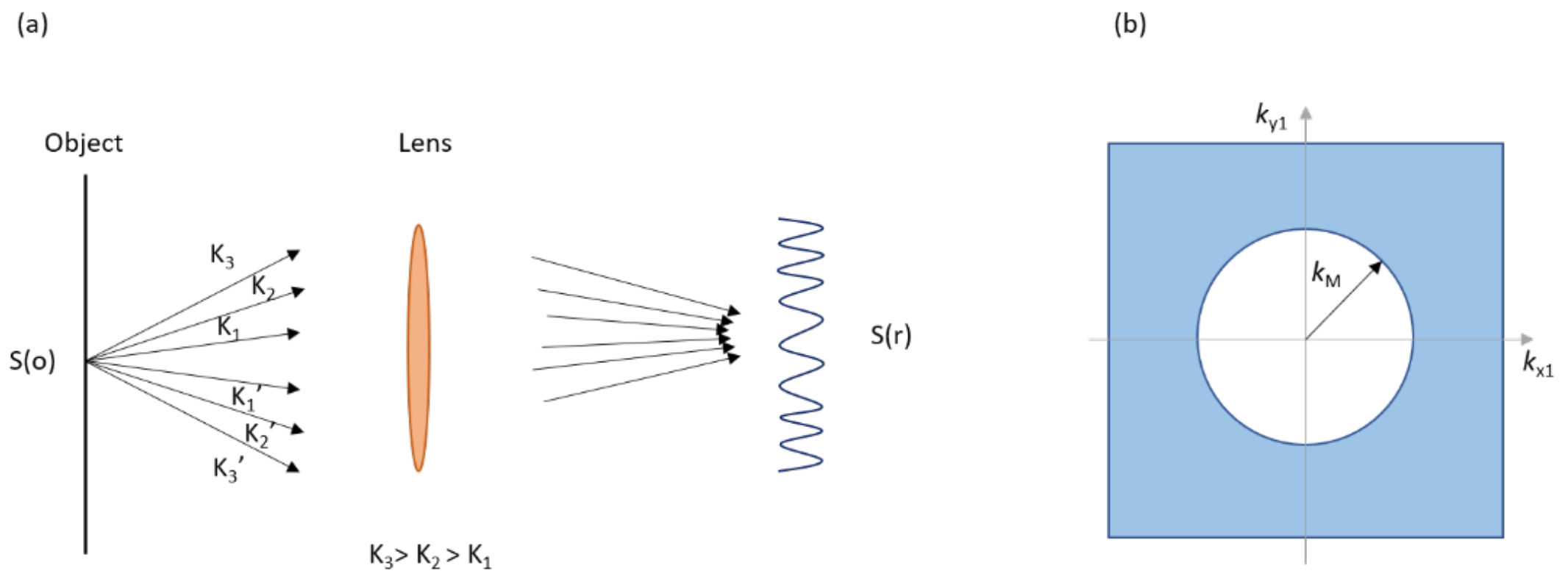

***Fig.S4.*** *(a) Abbe's picture of image formation, (b) Maximum frequency allowed by the aperture will define the resolution of the system.*

Now according to FT based approach, if $f(x)$ is the slit or object function (single or periodic), then the Fraunhofer image is given by its absolute-FFT,

The Fourier transformation of $f(x)$ is given by

$$f(q) = \int_{-\infty}^{\infty} f(x) e^{-2\pi i x.q} dx \tag{18}$$

The absolute of $f(q)$ is calculated by evaluating the Fourier integral over all frequencies and then taking square modulus of both sine and cosine parts separately for each frequency i.e. $q = -n \; to \; +n.$

$$Re(q) = \int_{-\infty}^{\infty} f(x) cos(2\pi x. q) dx \tag{19}$$

and,

$$Im(q) = \int_{-\infty}^{\infty} f(x) sin(-2\pi x. q) dx \tag{20}$$

Absolute FT of $f(x)$ for a given frequency $q$ is given by

$$Abs\ FT\ of\ f(x) = \sqrt{Re^2 + Im^2} \tag{21}$$

$$and\ the\ phase = \tan^{-1}\frac{Im}{Re} \tag{22}$$

The absolute FT is then plotted for each frequency and the result is the well-known Fraunhofer pattern.

The above Fourier based formalism interprets that the object function $f(x)$ acts as amplitude to expand into plane wave basis with different frequencies. Then each wave for a given frequency is integrated over certain $x$ range to obtain **Re** and **Im** part of the wave for each frequency. Finally, the amplitude (abs-FT) and phase of each Fourier waves is given by Eq. 10.

If we are interested in the object information, for the atomic systems it is the atomic potential and periodic form of this. This can directly be retried by inverse Fourier transform if $Re\ (x,k)$ and $Im(x,k)$ parts are known. However, these are not known in the experimental diffraction pattern, but is preserved if the intensity pattern is recorded in the image plane at atomic resolution. This implies that the phase information is lost only in the diffraction plane but not in the image plane.

❖ Thus, one can see that Fourier methods works by expanding the object function into plane waves of various frequencies. The phase can be calculated by $\tan^{-1}\frac{Im(k\ or\ x)}{Re(k\ or\ x)}$ in both diffraction plane or image plane. This phase value will depend on $f(x)$, its spatial distribution and magnitude. More on FT method on experimental high-resolution holography see sec. S2.2.

The analytical formula following Fraunhofer integral for single and periodic one-dimensional slit function are given below.

*For single slit:*

$$I(\theta) = I_0 \left[\frac{\sin\left(\frac{\pi a}{\lambda} sin\theta\right)}{\frac{\pi a}{\lambda} sin\theta}\right]^2 \tag{23}$$

*For N slits:*

$$I(\theta) = I_0 \left[\frac{\sin\left(\frac{\pi a}{\lambda} sin\theta\right)}{\frac{\pi a}{\lambda} sin\theta}\right]^2 \left[\frac{\sin\left(\frac{N\pi d}{\lambda} sin\theta\right)}{\sin\left(\frac{\pi d}{\lambda} sin\theta\right)}\right]^2 \tag{24}$$

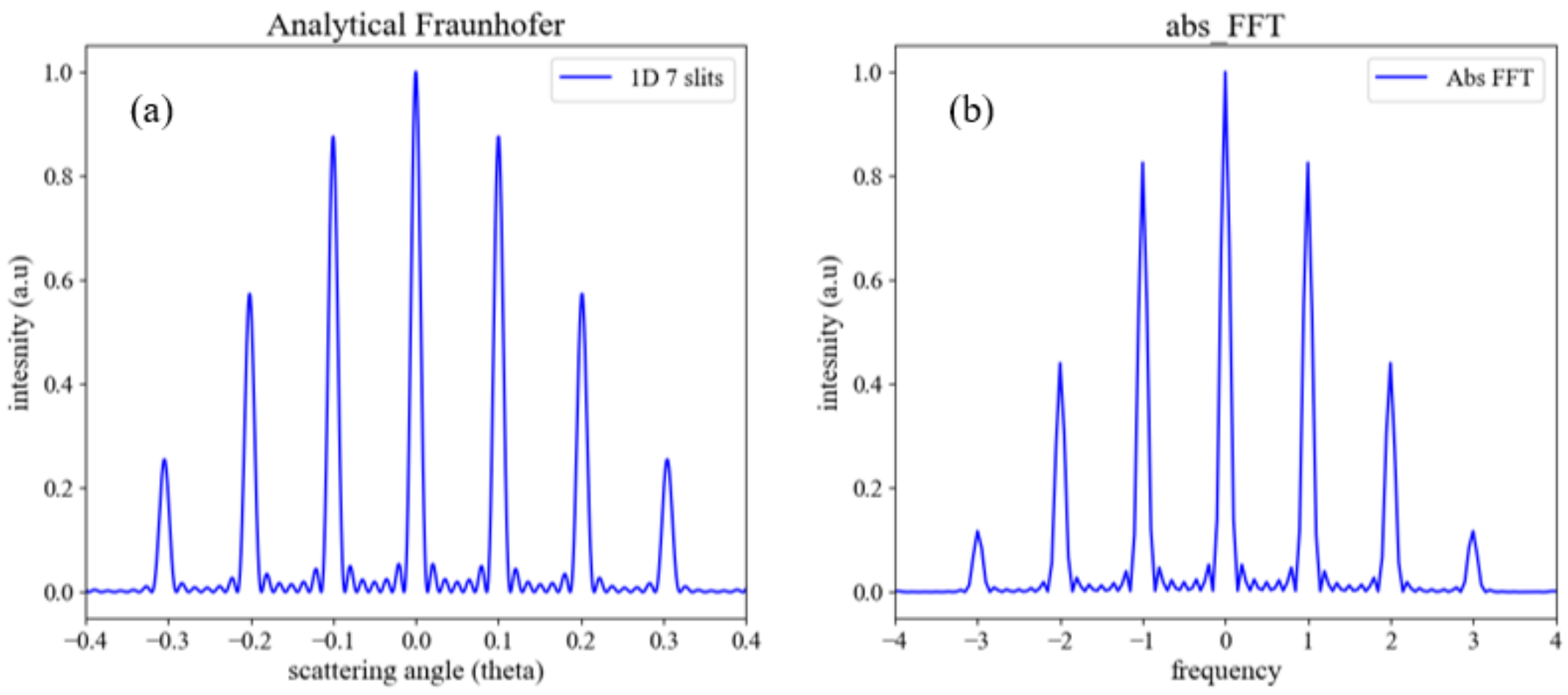

***Fig. S5.*** *(a) comparison of Fraunhofer pattern calculated by Fraunhofer analytical methods for periodic 7 slits with slit size a = 2 λ and periodicity d =10 λ and (b) Fourier transformation of 7 periodic slits followed by abs-FT. (b) appearance of zero magnitude in abs FFT is due to zero values of integration of* $\boldsymbol{Re(x,k)}$ *and* $\boldsymbol{Im(x,k)}$ *parts with respect to x corresponding to respective frequency values.*

## 1.4.Calculating intensity pattern for the atomic system

### 1.4.1. Using atom scattering factor

This is the method to calculate intensity pattern both at the far field and image plane. Fourier transformation is used to interchange between the plane. This is possible as the formalism of atom scattering factor retains the information on the scattering potential or in other words complete phase information.

The object exit wave function is given by the superposition between illumination plane wave and the scattered spherical waves with amplitude factor given by the atom scattering factor. The transmitted wave function in this case is derived from the Schrödinger integral equation and has the following form[2,4]

$$\psi_t(x) = \exp(2\pi i k_z z) + f_e(q)\frac{\exp\,(2\pi i q.r)}{r} \tag{25}$$

Where, $q = k - k_0$, and $f_e(q)$ is the atom scattering factor and is given by

$$f(q) = -\frac{m}{2\pi\hbar^2}\int V(r')e^{-2\pi i q.r}\,d^3r \tag{26}$$

The solution of wave function based on differential form of Schrödinger equation has the form of a plane wave. On the other hand, integral form gives solution of spherical waves along with amplitude factor as atom scattering factor. The equivalence between the two solutions is given in terms of envelop of all the spherical waves from many adjacent scattering centers will eventually result in a plane wave front. The picture is akin to Huygens's construction that a plane wave front is the envelope of many forward scattered spherical wavelets. This result is used along with the scattering factor as derived by Moliere to calculate the image of isolated atoms according to Huygens's principle at Fraunhofer approximation by using the following equation.[4,6]

$$g(x) = \left|1 + 2\pi i \int_0^{k_{max}} f_e(k) \exp[-i\chi(k)] J_0(2\pi kr) kdk\right|^2 \tag{27}$$

Where, $f_e(k)$ is the electron scattering factor in the Moliere approximation which has the advantage over Born scattering factor due to presence of imaginary part. $\chi(k)$ is the aberration function, $k_max = \alpha_{max}/\lambda$ (unit rad Å$^{-1}$) is the maximum spatial frequency in the objective aperture and $J_0(x)$ is the Bessel function of order zero, arising due to azimuthal integration of scattering factor.

❖ Therefore, from the Eq. S16, one can calculate image wave function amplitude and phase as a function of spatial coordinate and match with the reconstructed image of object exit wave function to interpret object structure.[7]

### 1.4.2. Weak phase object approximation (WPOA) for calculating intensity pattern

This is the widely discussed approach to explain phase contrast imaging in electron microscopy along with the consideration of lens aberration. The transmitted wave function without any lens aberration is given by

$$\psi_t(x) \sim t(x) \exp(2\pi i k_z z) \tag{28}$$

There is a close resemblance between Gabor's reading component of in-line hologram and the above expression. [see notes below marked with bullet].

The transmission function is given by

$$t(x) = \exp\,[i\sigma v_z(x)] \tag{29}$$

Now within weak phase object approximation (WPOA) series approximation yields

$$t(x) = 1 - i\sigma V_t(x, y) \tag{30}$$

Eq. 19 is similar to the Zernike phase object. After the lens response function acts on this, the image wave function is

$$\psi_i(x,y) = 1 - i\sigma\phi_p(-x,y) * \mathcal{F}\{P(u,v)\exp\big(i\chi(u,v)\big)\} \quad (31)$$

And the image intensity

$$I(x,y) = \psi_i(x,y)\psi_i^*(x,y) \approx 1 + 2\sigma\phi_p(-x,y) * \mathcal{F}\{sin\chi(u,v)P(u,v)\} \quad (32)$$

Where, $\phi_p(-x,y)$ and $V_t(x,y)$ have the same meaning. They are different as adopted from two different source.

According to Spence,[8] the image contrast is proportional to the projected specimen potential, convoluted with the impulse response of the instrument. In the absence of astigmatism, the radially symmetric point spread function (*psf*) or impulse response function can be written as

$$\sigma\mathcal{F}\{sin\chi(u,v)P(u,v)\} = \frac{2\pi}{\lambda^2}\int_0^{\theta_{ap}} sin\,\chi(\theta)J_0\left(\frac{2\pi\theta r}{\lambda}\right)\theta d\theta \quad (33)$$

The above intensity expression in Eq. S21 only in terms of sine convolution is obtained by omitting $\sigma^2$ term which associate cosine function [Ref. 9, page 487].[9]

Note that the PCTF act as envelope function to filter out certain frequency range in the diffraction plane. Aberration phase shift cannot be added with the phase term of diffracted wave or scattering factor,[6] this will mean shifting wave and consequently change in lattice parameter and this is never observed in the scattering plane. In real space the *psf* act as follows, the integration of $\sin\chi(\theta)$ over spatial frequency will give weight to the intensity due to pure phase transfer depending on the shape of the function, whereas the $J_0\left(\frac{2\pi\theta r}{\lambda}\right)$ will determine the resolution in terms of FWHM of the pattern. However, as the convolution procedure increase the volume of the resultant graph this needs to be normalized based on flux balance to see the effect of aberration in terms of drop in final intensity and broadening (see Fig. S7).

- ✓ **Analogy between Gabor's inline holography and HRTEM**

There is a close relationship between the mathematical formalism describing Gabor's inline holography and HRTEM image formation as described within the weak phase object approximation (WPOA).

The hologram is typically an interference pattern formed between incident plane wave illumination and scattered or diffracted wave due to object [Fig. S6 (a)]. It is the intensity distribution at some propagation distance, typically in the Fresnel regime. Holography allows for the interference effect to take place between the background wave and the scattered waves. Hologram is useful to reconstruct the image of the object faithfully after removing the geometric aberrations of the lens through subsequent ex-situ procedures.

The amplitude of the interference pattern is given by

$$A = \sqrt{UU^*} = \sqrt{A^{(i)2} + A^{(s)2} + 2A^{(i)}A^{(s)}\cos\left(\psi_s - \psi_i\right)} \qquad (34)$$

Where, $S$ is the source, $\sigma$ is a semitransparent object, $H$ is the screen, $U = Ae^{i\psi}$ is complex disturbance at a point in $H$, $A$ is the amplitude, $\psi$ is the phase, superscript and subscript corresponding to *(i)* and *(s)* denote incident and scattered waves, respectively.

Now, the reconstruction of the hologram is written in terms of multiplication of object transmission function with the plane wave illumination (replacing wave) and together they form the transmitted wave. Thus, the reconstructed wave or substituted wave is written as follows [Fig. S6 (b)]

$$U' = \alpha_p U^{(i)} = KA^{(i)2}e^{i\psi_i}[A^{(i)} + \frac{A^{(s)2}}{A^{(i)2}} + A^{(s)}e^{i(\psi_s-\psi_i)} + A^{(s)}e^{-i(\psi_s-\psi_i)}] \quad (35)$$

for $\Gamma = 2$. Where $\alpha_p$ is the amplitude transmission factor of the positive, and for details on other parameters, see Ref [3].

As can be recalled that the transmitted wave function in HRTEM within weak phase object approximation (WPOA) has the form [4]

$$\psi_t(x) \sim t(x) \exp(2\pi i k_z z) \quad (36)$$

The similarity can be noted between Eq. (24) and (25).

In the transmitted wavefunction, the information about the object is carried by the transmission function. The formation of image directly from the object using appropriate lens is equivalent to form the same image from the recorded hologram with a supplementary illumination with the help of a lens [Fig. S6 (c)].

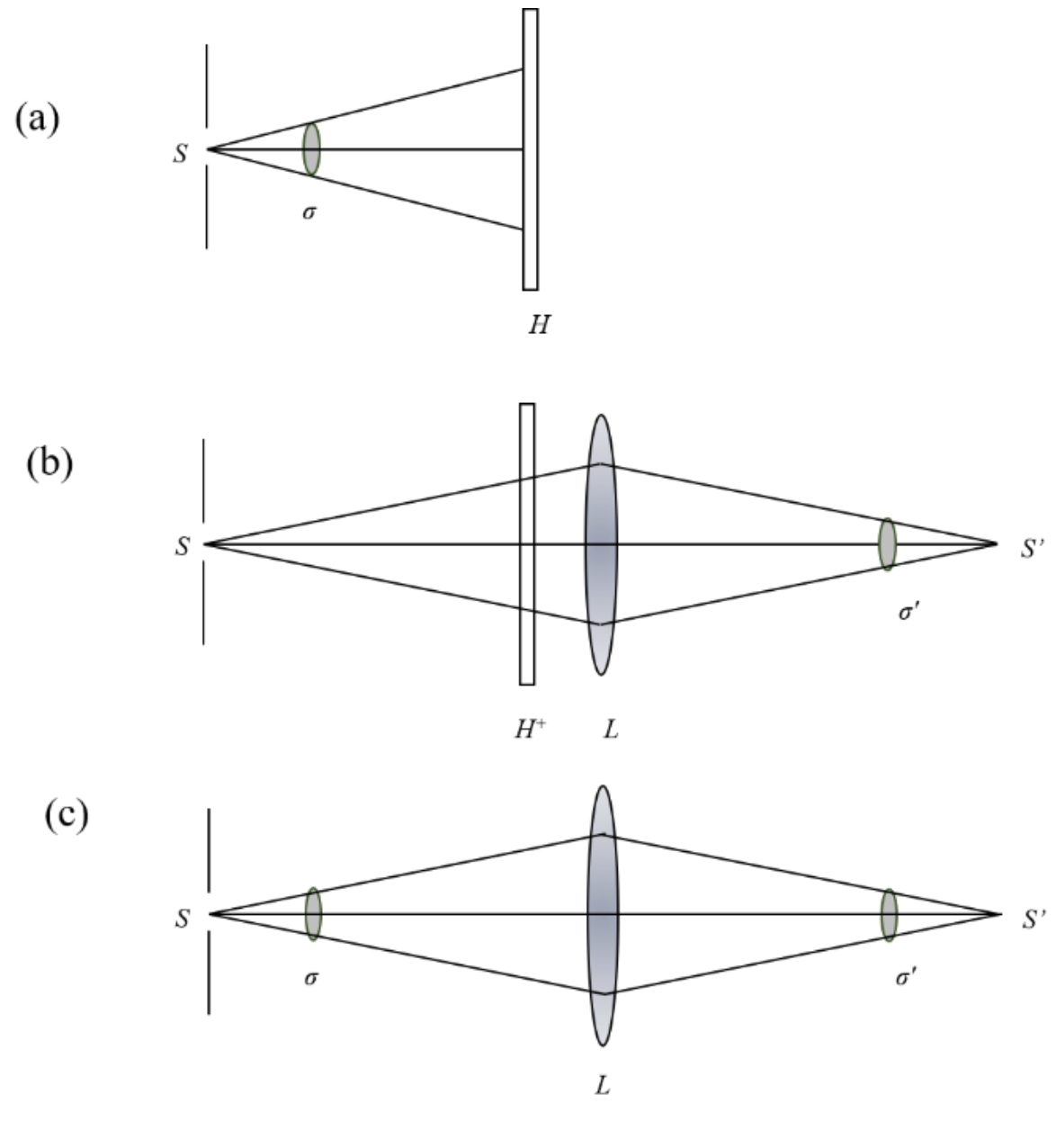

***Fig. S6.*** *(a) & (b) Schematics showing the formation of hologram and image of the object as reconstructed from the recorded hologram. (c) Equivalent image formation without*

*hologram. Hologram helps to removes the lens defect by ex-situ processes to faithfully form the image of the object.*

✓ **Note on series approximation in Eq. S19**

In the transmitted wavefunction, the information about the object is carried by the transmission function (see standard definition in the following note). The quantities inside the exponential of the transmission function is the change in phase and can be added with the plane wave phase of the illumination as follows;

$$\psi_t(x) \sim t(x)\exp(2\pi i k_z z) \tag{37}$$

$$\psi_t(x) \sim \exp\big(i(2\pi k_z z + \sigma v_z(x)\big) \tag{38}$$

To read this additional phase shift we need to have a reference wave with respect to that the fringe shift will be visible and change in phase can be measured. However, in HRTEM this missing information is determined by series approximation within WPOA and with the help of aberrations and focus settings which act as phase plate to retrieve the missing information.

In Fourier space, following the Scherzer's argument that the additional phase shift due to aberration act as frequency filter envelope and the effect is introduced by convolution in real space.[6] However, we show below that if one adds any additional phase shift with the original plane wave phase then this information is lost completely.

Now, if we do not do any such series approximation in Eq. S19 then the phase information will be inside the cosine and sine trigonometric functions and intensity of transmitted radiation derived by multiplying the transmitted wave function with its complex conjugate will result in a constant value and thus phase information is lost. However, one can notice that by expanding the transmission function in a series (weak phase object approximation) the phase is recovered and multiplying with complex conjugate will retain the phase. Thus, we can say that by mere mathematical manipulation we are recovering phase which should not be the case.

Therefore, without series expansion and approximation then,

The image wave function is

$$\psi_i(x,y) = \{t(x)\exp(2\pi i k_z z)\} * \mathcal{F}\{P(u,v)\exp\big(i\chi(u,v)\big)\} \tag{39}$$

Image intensity

$$I(x,y) = 1 \tag{40}$$

See derivation below

$$p = 2\pi k_z z + \sigma v_z(x) \tag{41}$$

$$\psi_i(x,y) = \{\cos(p) + isin(p)\} * \{\cos(x) + isin(x)\} \tag{42}$$

$$I(x,y) = \psi_i(x,y)\psi_i^*(x,y) \approx 1 \tag{43}$$

Thus, the phase can only be recovered under mathematical series expansion (weak phase object approximation).

✓ **Note on transmission function and phase object:**

The definition of a transmission function can be found in [Ref. 3, p 446-447].[3] Any diffraction grating which introduces variation in amplitude and phase on the incident wave can be characterized by its transmission function. It is given by

$$F(\xi,\eta) = \frac{V(\xi,\eta)}{V_0(\xi,\eta)} \quad (44)$$

Where, $V_0(\xi,\eta)$ is the disturbance on $(\xi,\eta)$ plane in the absence of object and $V(\xi,\eta)$ is the disturbance on the same plane when object is present. This is consistent with the formalism of transmission function for electron imaging discussed before.

In page 472 of Born and Wolf, in Zernike's phase contrast method section, the phase object is defined by a complex amplitude function (for light) as follows

$$F(x) = e^{i\phi(x)} \quad (45)$$

Where, $\phi(x)$ is a real periodic function and whose period is equal to the period of grating [in case of periodic grating]. For $\phi(x)$ small compared to unity the above equation can be approximated to

$$F(x) \sim 1 + i\phi(x) \quad (46)$$

The effect of Zernike phase plate modifies the intensity of the object wave that depends linearly on the phase change due to object i.e., $I(x) = 1 \pm 2\phi(x)$.

Now one can notice the origin of observing $\phi(x)$ in the intensity information. If this approximation is not done, $\phi(x)$ would be lost. This is similar to the discussion already made in the context of WPOA and transmission function.

✓ **Transmission function in HRTEM**

In case of HRTEM, the transmission function has the origin in change in wave vector of electron after interacting with the specimen potential.[4,9]

The wave function of plane wave travelling along the *z* direction of optic axis is given by

$$\psi(x) = \exp(2\pi i k_z z) = \exp\left(\frac{2\pi i z}{\lambda}\right) \qquad (47)$$

The quantum mechanical Planck constant enter into the expression through relativistic expression of electron wave vector in vacuum.

$$k_z = \frac{1}{\lambda} = \frac{\sqrt{eV(2m_0c^2+eV)}}{hc} \qquad (48)$$

The wavevector of electron in the specimen is given by

$$\frac{1}{\lambda_s} = \frac{[(eV+eV_s)(2m_0c^2+eV+eV_s)]^{1/2}}{hc} \qquad (49)$$

Kindly note that the above expression changes the wave vector of vacuum electron.

After series expansion and keeping only lowest order terms in $\frac{V_s}{V}$, the transmitted wave function yields

$$\psi_t(x) \sim \exp(2\pi i k_z z)\exp(i\sigma V_s z) \qquad (50)$$

$$\sigma = \frac{2\pi}{\lambda V}\left(\frac{m_0c^2+eV}{2m_0c^2+eV}\right) = \frac{2\pi m e\lambda}{h^2} \qquad (51)$$

The transmitted wave function can now be written as

$$\psi_t(x) \sim t(x)\exp(2\pi i k_z z) \qquad (52)$$

Where, $t(x)$ is called transmission function and is given by

$$t(x) = \exp[i\sigma v_z(x)] \qquad (53)$$

- ✓ **Note on convolution:**

Below is the effect of convolution operation on a model function.

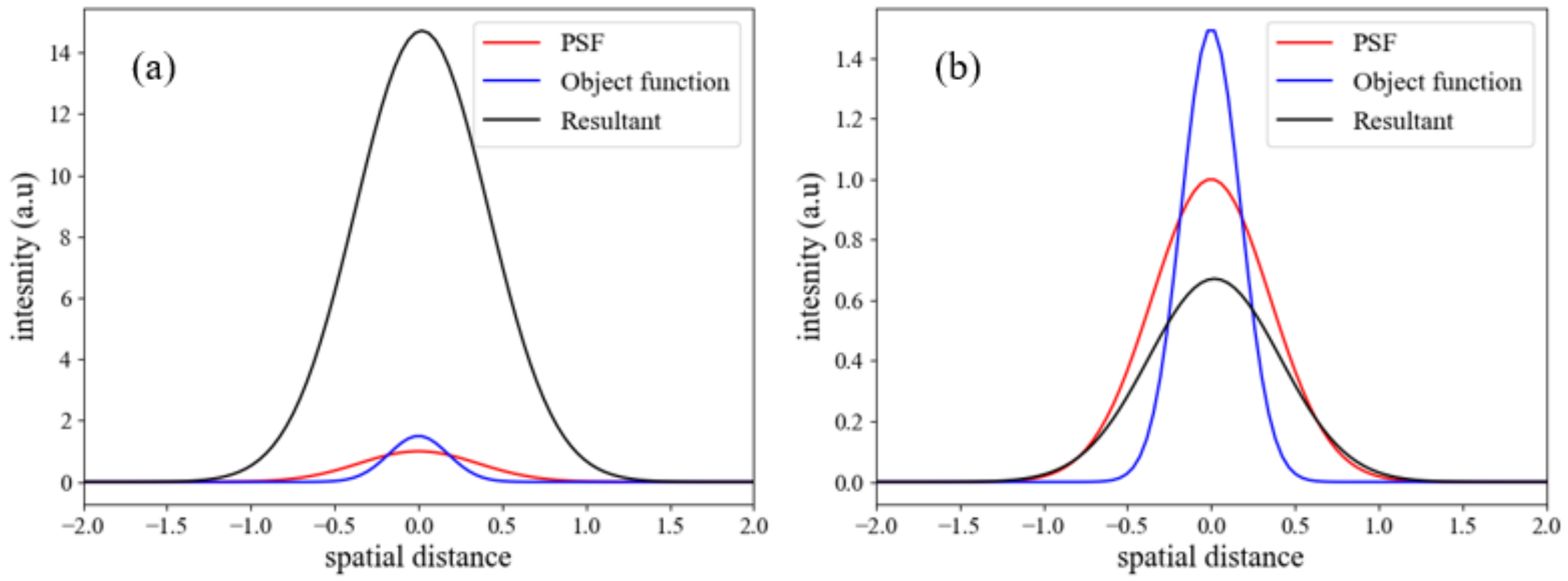

***Fig. S7.*** *(a) Examples showing the area under the graph (resultant black) is not preserved after convolution between a model object function (Gaussian blue, with peak value 1.5 and FWHM 0.5) and convoluting function (Gaussian red, with peak value 1and FWHM 1). (b) Resultant black is normalized based on total area under the curve of model object function (flux balance) which then shows the spread at the expense of reduced peak value.*

This convolution operation changes the peak height of the resultant curve significantly. Therefore, one need to normalize the resultant based on flux balance i.e. area under the model object function. After considering flux balance the resultant curve drops in peak value compared to the object function.  Experimentally acquired images need to be deconvoluted with the known *psf* and compare the image with the simulated image.

## 2. Various schemes on HRTEM focal series reconstruction

### 2.1. Description of focal variation methods according to Saxton

Saxton showed the equivalence between different reconstruction techniques in terms of equivalence in restoring filter.[10] The effect of restoring filter is equivalent to deconvolving the effect of aberration and removing other unwanted components. The effect of coherence was not considered in the restoration filter used in the 3D paraboloid approach. But Wiener approach consider the coherence effect which is the actual resolution limiting factor. However, after appropriate approximations the form of restoring filter becomes similar to the incoherent case. The simplified form of the restoring filter is given by

$$r_n(k) = \frac{1}{N} \exp\{i\gamma(k)\} \quad (54)$$

And the aberration function

$$\gamma_n(k) = \pi C_s \lambda^3 k^4 - \pi \lambda z_n k^2 \quad (55)$$

Saxton also derived the final form of the wave function which can be retrieved through restoration filter as proposed by van Dyck in the 3D paraboloid method and later connected with the Schiske's earlier work on this, from a different start in terms of the form of intensity expression and without considering any restoration filter.[5,11,12] This was based on first taking several images at different focus and then adding the waves in reciprocal space after multiplying with the phase conjugate corresponding to each defocus. This approach provides insight on the effect of restoration on the conjugate wave function and non-linear image component. Now, to follow the procedure, under perfect coherence the image intensity at focus *z* was written as

$$i(x,z) = |1 + \psi_i|^2 = 1 + \psi_i(x,z) + \psi_i^*(x,z) + h(x,z) \quad (56)$$

With $h = |\psi_i|^2$ is the unwanted non-linear term, $\psi_i(x,z)$ is the scattered part of the wave,

The 2D Fourier transformation of above equation and then adding explicitly dependence on defocus results in the following expression.

$$I(k, z) = \delta(k) + \psi(k)\exp(\pi i\lambda zk^2) + \psi^*(-k)\exp(-\pi i\lambda zk^2) + H(k, z) \quad (57)$$

Where, $H(k, z)$ is the non-linear component that describes the autocorrelation between two linear terms.

Now, by assuming $k$ is nonzero which allows to omit delta function and constant $k$, the image at some focus $z_n$ can be written as

$$I(z_n) = I_n = \psi\exp(-\pi i\lambda z_n k^2) + \psi^*\exp(-2\pi i\lambda z_n k^2) + H_n \quad (58)$$

Now, within coherent detection i.e., multiplying by a phase conjugate and summing over N, the image intensity becomes

$$\sum_n I_n \exp(-\pi i\lambda z_n k^2) = \psi\sum_n 1 + \psi^*\sum_n \exp(-2\pi i\lambda z_n k^2) + \sum_n H_n \exp(-\pi i\lambda z_n k^2) \quad (59)$$

From the above expression, the wave function $\psi$ thus accumulates to N times to its original value. The sum over $\psi^*$ accumulates to $\sqrt{N}$ times to its original value. The non-linear term $H_n$ is considered to vary randomly with defocus. Thus, by dividing the above equation by N, the wave function may be recovered. This is the expression which is used generally in all through focus series method-based wave function retrieval.

- ✓ **Note on Eq. (46)**

In the above discussion the effect of Cs is ignored but can be incorporated through the exponential phase function or transfer function. Experimentally this is akin to applying a frequency filter formed by known complex transfer function corresponding to every focus value to the Fourier space intensity and then summing all of them to retrieve the most dominant wave function compared to other contributions, namely conjugate part and non-linear part. The complex quantity thus obtained can be used to calculate the amplitude and phase part of the object exit wave (OEW) function by arctan function.

In fact, the Eq S45 can also be written in following way following the description on holography as given in Born and Wolf [p 504, section 8.10.1].[3] in Eq. S45, the illuminating wave function $\psi_0$ considered to be 1 similar to Schiske's axial illumination condition for which $k_0 = 0$.

$$i(x,z) = |\psi_0 + \psi_i|^2 = \psi_0\psi_0^* + \psi_0^*\psi_i(x,z) + \psi_0\psi_i^*(x,z) + \psi_i(x,z)\psi_i^*(x,z)$$

$$= A_0^2 + A_i^2 + A_0A_i \exp i\,(\phi_i - \phi_0) + A_0A_i \exp i\,(\phi_0 - \phi_i)$$

$$= A_0^2 + A_i^2 + 2A_0A_i cos(\phi_i - \phi_0) \qquad (60)$$

The one below after following Born and Wolf directly

$$= |\psi_0 + \psi_i|^2 = |A_0 \exp(i\phi_0) + A_i \exp(i\phi_i)|^2$$

$$= \exp(i\phi_0) \exp(-i\phi_0)\,|A_0 + A_i \exp i(\phi_i - \phi_0)|^2$$

$$= A_0^2 + A_i^2 + 2A_0A_i cos(\phi_i - \phi_0) \qquad (61)$$

Two slightly different starting arrangement give the same results. The quantities are all real and different than expression in Eq. S45. Now it depends on the vector nature of the momentum direction whether we will have in-line or off-axis like geometry. In case of off-axis geometry another phase term appears, component of wave vectors $Qx$ inside the cosine term, which gives a measure of carrier frequency and a reference point for a fringe shift.

Moreover, the second line of the expression Eq. S49 is in fact in an intermediate state, where imaginary part will cancel each other automatically and yield only real part. This means that by fitting the intensity equation and evaluating cosine term the relative phase change in the image can be extracted.

### 2.2. Reconstruction scheme in off-axis electron holography

However, in the standard formalism of off-axis electron holography, one of the two side bands (SBs) is used to perform inverse Fourier transformation to recover the wave function. The SBs are the convolution between the FT of wave function and the delta function. The delta function ensures shifting of the FT of wave functions in frequency space, thus separates it from the central band (CB) and conjugate wave function part.

According to a lecture presentation slide by James Loudon,[13] for more details see Ref.[14,15] the geometry of interference in off-axis holography is given in Fig. S8.

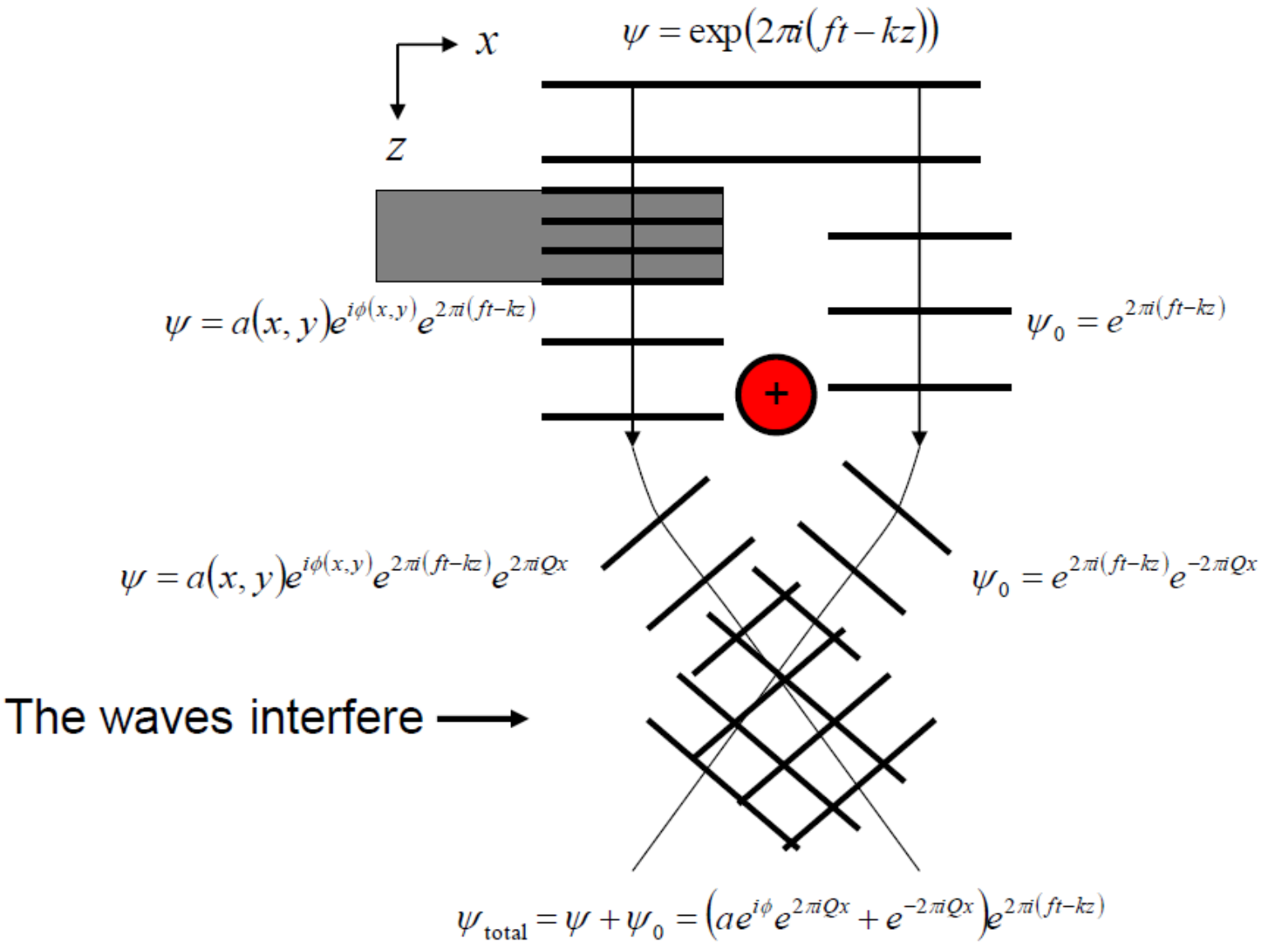

***Fig.S8.*** *Schematic of transmission Off-axis electron holography interference geometry. The probe plane wave split into two halves are brought together for interference. One of the two partial waves goes through sample and another through vacuum serving as a reference wave. The sample induces additional phase shift which can be used to interpret the object structure.*

$$\psi_{total} = e^{2\pi i(ft-kz)}\left(ae^{i\phi}e^{2\pi iQx} + e^{-2\pi iQx}\right) = e^{2\pi i(ft-kz)}e^{i\frac{\phi}{2}}(ae^{i\frac{\phi}{2}}e^{2\pi iQx} + e^{-i\frac{\phi}{2}}e^{-2\pi iQx}) \tag{62}$$

$$I_{total} = |\psi_{total}|^2 = \psi^*_{total}\psi_{total} = (ae^{-i\frac{\phi}{2}}e^{-2\pi iQx} + e^{i\frac{\phi}{2}}e^{2\pi iQx})(ae^{i\frac{\phi}{2}}e^{2\pi iQx} + e^{-i\frac{\phi}{2}}e^{-2\pi iQx}) \tag{63}$$

$$I_{total} = 1 + a^2(x,y) + 2a(x,y)\cos\left(4\pi Qx + \phi(x,y)\right) \tag{64}$$

$$\begin{aligned} I(x) &= 1 + a^2(x) + 2a(x)\cos\left(4\pi Qx + \phi(x)\right) \\ &= 1 + a^2(x) + a(x)e^{i\phi(x)}e^{4\pi iQx} + a(x)e^{-i\phi(x)}e^{-4\pi iQx} \end{aligned} \tag{65}$$

The Eq. S54 according Ref. 15 is

$$I_{hol}(\vec{r}) = I_0 + I_{ima,inel}(\vec{r}) + I_{ima,el}(\vec{r}) + 2|\mu|A_0A_{el}(\vec{r})\cos\left(2\pi\vec{q}_c\vec{r} + \phi(\vec{r})\right) \tag{66}$$

The intensity expression in Eq. S54 is Fourier transformed and using $\delta(q-Q) = \int_{-\infty}^{\infty} e^{2\pi i(q-Q)x}dx$ that separates the DC part and two twin image components (SBs) in the frequency plane.

$$F.T.[I(x)] = \delta(q) + F.T.[a^2(x)] * \delta(q) + F.T.\left[a(x)e^{i\phi(x)}\right] * F.T.\left[e^{4\pi iQx}\right] + F.T.\left[a(x)e^{-i\phi(x)}\right] * F.T.\left[e^{-4\pi iQx}\right]$$

$$F.T.[I(x)] = \delta(q) + F.T.[a^2(x)] * \delta(q) + F.T.\left[a(x)e^{i\phi(x)}\right] * \delta(q-2Q)\phi + F.T.\left[a(x)e^{-i\phi(x)}\right] * \delta(q+2Q)$$

$$F.T.[I(x)] = \delta(0) + F.T.[a^2(x)] * \delta(0) + F.T.\left[a(x)e^{i\phi(x)}\right] * \delta(q-2Q) + F.T.\left[a(x)e^{-i\phi(x)}\right] * \delta(q+2Q)$$

(67)

The Eq. S56 according to Ref. 15 is

$$FT[I_{hol}(\vec{r})] = I_0\delta(\vec{q}) + FFT\left[I_{ima,el}(\vec{r}) + I_{ima,inel}(\vec{r})\right] \otimes \delta(\vec{q}) + \hat{V}FFT\left[A_0A_{el}(\vec{r})e^{i\phi(\vec{r})}\right] \otimes \delta(\vec{q}-\vec{q}_c) + \hat{V}FFT\left[A_0A_{el}(\vec{r})e^{-i\phi(\vec{r})}\right] \otimes \delta(\vec{q}+\vec{q}_c)$$

(68)

Now selecting one of the two SBs → $F.T.\left[a(x)e^{i\phi(x)}\right]$

Inverse FT of the sideband → $\left[a(x)e^{i\phi(x)}\right]$ gives the wave function.

The spatial resolution of the technique is determined by the size of the mask placed around the sideband.

Now there is a fundamental difference in the intensity expression as given by Born and Wolf on Gabor's in line holography and the expression corresponding to off-axis electron holography. In off-axis geometry an additional phase term in form of wave vector component $Q = \vec{q}_c = 2k_x$ appears due to wave interference at an angle. This defines the carrier frequency in the off-axis electron hologram and changes with the angle of interference which can be controlled by the biprism voltage. Larger the angle of interference, larger will be the component of horizontal wave vector and finer the fringe spacing. Information about the object phase is carried through $\phi(x)$ term and appears as a shift on the hologram fringes. In case of inline holography, it is the $(\phi_i - \phi_0)$ which appears inside the cosine function and only the relative phase change in terms of intensity between two object points can be observed in the intensity pattern.

Now one can also find a similarity of the intermediate state of the expression in the form of wave functions between Eq. S50 with that of Eq. S54 of off-axis holography.

- ❖ Now Eq. S54 can be understood in comparison to Eq. S49 & Eq. 8 as follows. We also describe importance of some components in the equations and what they describe and do not describe and some minor issues with the description in terms of amplitude term $a^2(x)$ and $Q$ in the Eq. S54 &56. There is a difference between the Eq. S54 & S56 describing holographic phase shift or fringe bending in terms of $\phi(x)$, separation of twin image wave functions and the way actual FT works on the experimental HRTEM image intensity embedded with the atomic resolution hologram i.e. $I(x)\ or\ I(x,y)$ as given in Eq. 8 and S49. What Eq. S54 is stating is that the object phase information $\phi(x)$ appears as fringe bending in the recorded hologram with carrier frequency $Q$ in the image plane. At atomic resolution, the terms $a^2(x)$ and $F.T.[a^2(x)]$ in the Eq. S53 or S56 do not describe the HRTEM intensity pattern like Eq.49 or 50, which has an exclusive cosinusoidal term $A_0^2 + A_i^2 + 2A_0A_icos(\phi_i - \phi_0)$ in HRTEM intensity expression with object phase term $\phi_i$ within it that through the interference pattern carry the information in the form of lattice structure and interpret the associated intensity in terms of object potential. Though it was mentioned appropriate way in Ref. 15 through $I_{ima,el}(\vec{r})$ [Eq. S 55], however, $I_{ima,el}(\vec{r})$ is written based on OEW of the form $A(\vec{r})e^{i\phi(\vec{r})}$ which is similar to the one used to derive Eq. S54. The factor $2A_0A_i$ outside the cosine term is a non-linear term and this cannot be accessed directly and can be used as a uniform back ground in the way described in the manuscript through Eq. 9 where the overall intensity value is used due to such term. Moreover, the amplitude of individual complex waves like $a(x)$ or $A(\vec{r})$ can only be read through $Re[a(x)\cos\phi_i]$ or $Im[a(x)\sin\phi_i]$ forming interference patterns and the intensity or modulus square corresponding to it in the form of $a^2(x)$ should not have any structure in terms of spatial variation in intensity and only contributes to the uniform background. Similarly, the modulus of individual wave function $\psi_0$ or $\psi_i$ also will not show ant structure and give only constant modulus term contributing to background. Therefore, it is only through the trigonometric functions upon interference between waves show up intensity pattern. However, Eq. S56 and S57 predicts a structure in spatial intensity through $\delta(q) + F.T.[a^2(x)] * \delta(q)$ to explain FFT patterns in CB which is not the true situation. Now the $F.T.\left[e^{4\pi iQx}\right]$ or $\delta(q-2Q)$ and $F.T.\left[e^{-4\pi iQx}\right]$ or $\delta(q+2Q)$ terms ensure displacement of $F.T.\left[a(x)e^{i\phi(x)}\right]$ term on the frequency axis having information on spatial modulation of intensity and without constant background term. Now the inverse FT of $\delta(q+2Q)$ (it should be $Q$ and not $2Q$, see Eq. S57 and further discussion latter) term gives uniform background in the image plane and inverse FT of $F.T.\left[e^{-4\pi iQx}\right]$ returns the information on carrier frequency and fringe bending with respect to object phase in the image plane if part of this phase term is added with $F.T.\left[a(x)e^{i\phi(x)}\right]$ term before performing inverse FT. Therefore, one can see that the particular way of using terms can yield different information upon inverse FT. Thus, the transfer of information on fringe bending is not captured in the FT pattern, it is only captured in the image plane!! It is the $F.T.\left[a(x)e^{i\phi(x)}\right]$ that contains the object phase information. In fact, the correct form of the object exit wave function (OEW) of the form $a(x)e^{i\phi(x)}$ used should have a relative phase term due to self-interference [see Eq.

S49] and absolute of which would show up copy of abs-FT pattern of HRTEM image intensity around the two SBs similar to CB and any abs-FT of HRTEM image intensity alone. We elaborate more and describe the difference below.

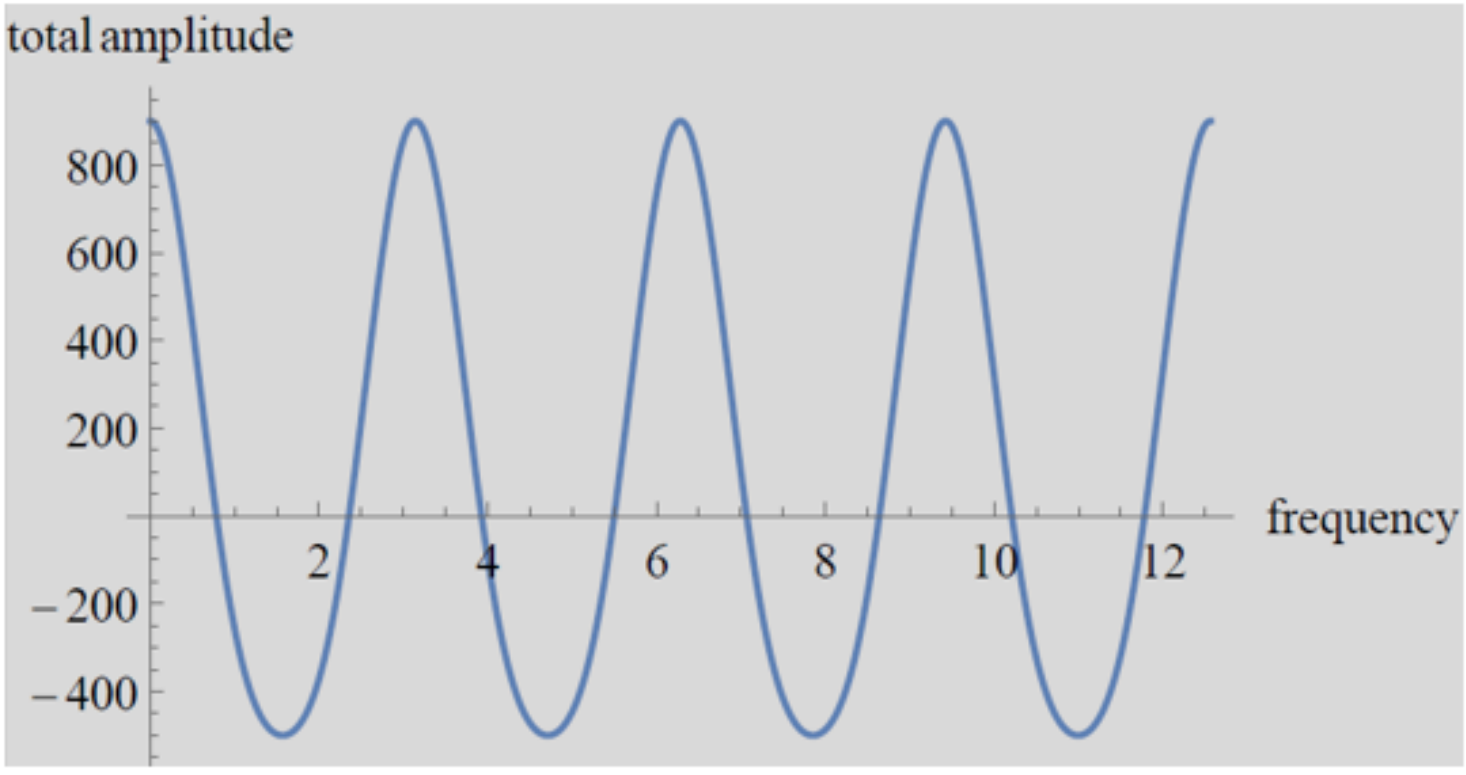

Plot[{(700 + 200 * Cos[2x]) * Cos[2x]}, {x, 0,4Pi}]

***Fig.S9.*** *Schematic showing multiplication of two cosine function having same frequency will result in non-zero integration value for a given period. This is similar to formation SB intensity while performing FT of HRTEM image embedded with hologram.*

In Fig. S8, the OEW due to one half of the wave should have a form as described in Eq. S49 or S50. Therefore, we modify the Eq. S54, S55, S56, S57 as follows, and present slightly different forms and finally compare it with the actual FT of HRTEM image with hologram.

$a(x)e^{i\phi(x)}$ becomes $A_0 \exp(i\phi_0) + A_i \exp(i\phi_i)$, we can call it $\psi_{inline}$ or $\psi_{HRTEM}$ and the intensity $I_{HRTEM} = A_0^2 + A_i^2 + 2A_0A_i cos(\phi_i - \phi_0)$, is the modulus square of $\psi_{HRTEM}$ as already given in Eq. S49 or S50. The $I_{ima,el}(\vec{r})$ term used in Eq. S55 &S57 should be considered as $I_{HRTEM}$.

$$\psi_{HRTEM} = \psi_{inline} = A_0 \exp(i\phi_0) + A_i \exp(i\phi_i) \qquad (69)$$

Thus, Eq. S51 modifies to

$$\psi_{total} = \left\{\left(A_0 e^{i\phi_0} + A_i e^{i\phi_i}\right) e^{2\pi i Q x} + A_0 e^{i\phi_0} e^{-2\pi i Q x}\right)\right\}$$

$$= \left\{\left(\psi_{HRTEM} e^{2\pi i Q x} + A_0 e^{i\phi_0} e^{-2\pi i Q x}\right)\right\} \quad (70)$$

And Eq. S52, modifies to

$$I_{total} = |\psi_{total}|^2 = \psi^*_{total}\psi_{total}$$

$$= \left(\psi_{HRTEM} e^{2\pi i Q x} + A_0 e^{i\phi_0} e^{-2\pi i Q x}\right)\left(\psi^*_{HRTEM} e^{-2\pi i Q x} + A_0 e^{-i\phi_0} e^{2\pi i Q x}\right)$$

$$(71)$$

$$I_{total} = A_0^2 + \psi_{HRTEM}\psi^*_{HRTEM} + A_0 e^{-i\phi_0}\psi_{HRTEM} e^{4\pi i Q x} + A_0 e^{i\phi_0}\psi^*_{HRTEM} e^{-4\pi i Q x}$$

$$= 1 + \{\, A_0^2 + A_i^2 + 2A_0 A_i cos(\phi_i - \phi_0)\} + \left(\, A_0^2 + A_0 A_i e^{i(\phi_i - \phi_0}\right) e^{4\pi i Q x}$$

$$+ \left(\, A_0^2 + A_0 A_i e^{-i(\phi_i - \phi_0}\right) e^{-4\pi i Q x}$$

$$(72)$$

And

$$F.T.(I_{total}) = F.T.(1) + F.T.(I_{HRTEM}) + F.T.\left(A_0 e^{-i\phi_0}\psi_{HRTEM}\right) * \delta(q - 2Q)$$

$$+ F.T.(A_0 e^{i\phi_0}\psi^*_{HRTEM}) * \delta(q + 2Q)$$

$$(73)$$

Eq. S62 can also be written as which is the correct form in terms of $\delta(q + Q)$, as used in Eq. 57 [Ref. 15].

$$F.T.(I_{total}) = F.T.(1) + F.T.(I_{HRTEM}) + F.T.\left(A_0 e^{-i\phi_0}\psi_{HRTEM} e^{2\pi i Q x}\right)$$

$$* \delta(q - Q) + F.T.(A_0 e^{i\phi_0}\psi^*_{HRTEM} e^{-2\pi i Q x}) * \delta(q + Q)$$

$$(74)$$

The Eq. S62 now correctly described the HRTEM image recorded with atomic resolution hologram in the CB, and the wavefunction $\psi_{HRTEM}$ has plane wave term $A_0 e^{-i\phi_0}$ outside of it. However, absolute square of it should now have a structure as observe around SB (see Fig. S10). Earlier consideration of the wavefunction in the form $a(x)e^{i\phi(x)}$ will not return any structure unless a cosine term is involved upon absolute square, this is only ensured by writing the $I_{HRTEM}$ in the form of Eq. S49 or S50.

Eq S63 suggests that while performing inverse FT cosine fringes will be available in the real space image and the location of SBs will be at Q and not 2Q. In fact, the position at Q is more correct as in the calibrated FFT pattern SBs appears at Q, i.e., ~ at 2.5 Å$^{-1}$ for holography fringe spacing of 0.4 Å [Fig. S10]. Though, the correct form of this is used in Ref. 15 but additional phase term $e^{2\pi iQx}$ is not considered. However, after absolute the terms $e^{-i\phi_0}$ and $e^{2\pi iQx}$ will be cancelled out and only $A_0{}^2 I_{HRTEM}$ will exist. This is in fact more realistic situation if one takes a close look on how the FT of the experimental HRTEM image including hologram works. Experimental hologram is a combination of (HRTEM pattern + cosinusoidal hologram) on the sample area and cosinusoidal hologram in the vacuum area with constant background for both type of areas.

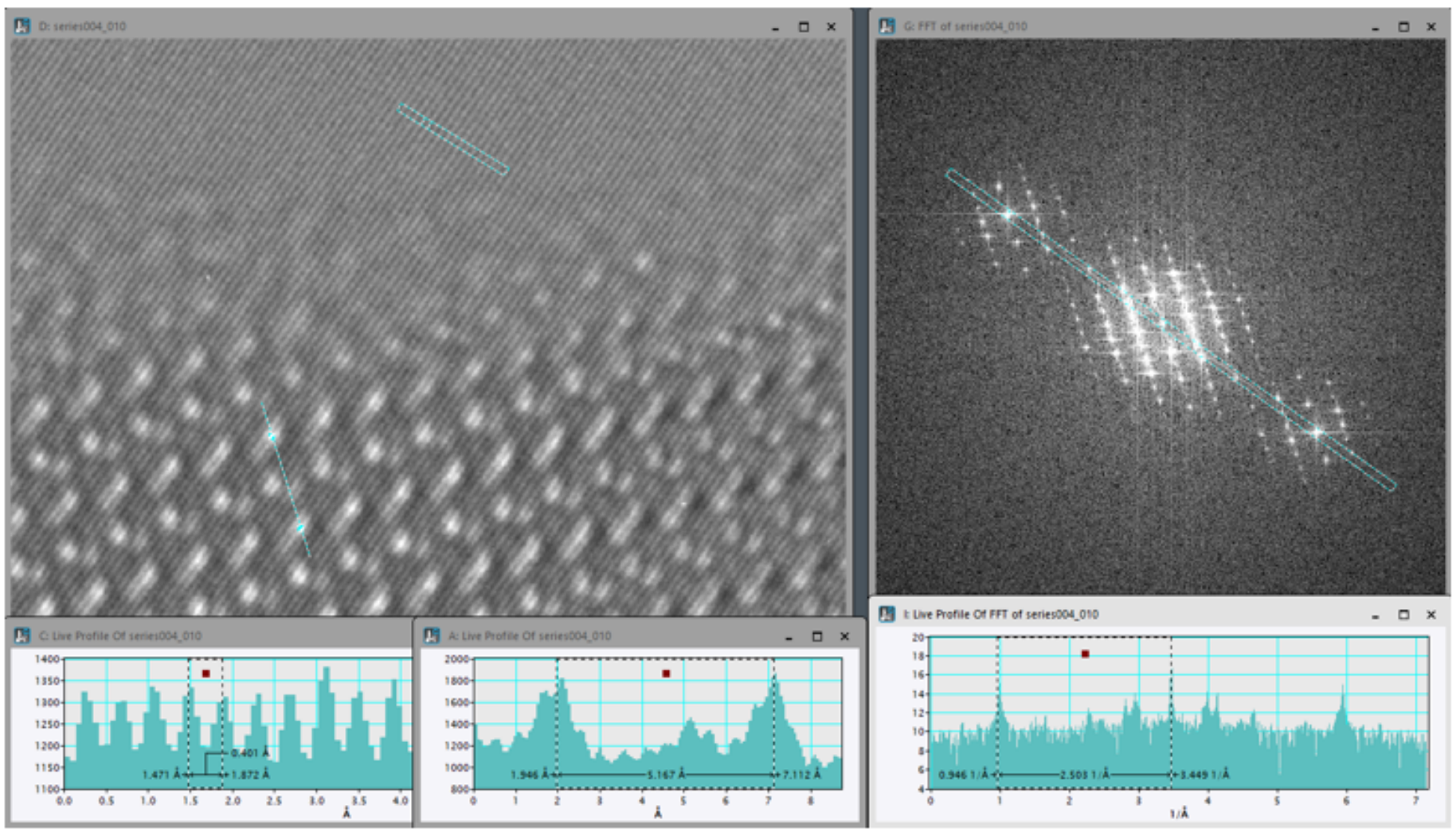

***Fig.S10.*** *ZnO lattice image with atomic resolution hologram with fringe periodicity of 40 pm. Side band location is at 1/40 pm in the absolute FT image on the right hand side.*

Now by taking a closer look on how the FT process (sec. S1.3) work on the experimental image we can understand the origin of CB and two SBs in off-axis electron holography and compare the origin of twin images between off-axis and in-line holography. To repeat again, the generation of abs-FFT pattern work as follows, for any function $f(x)$, in the present example a HRTEM image with atomic resolution hologram, will be expanded into various cosine and sine harmonics. It is the integration value of $f(x)\cos(2\pi kx)$ or $f(x)\sin(2\pi kx)$ for a given spatial range will determine the pattern. Presence of intensity in a FT spot means the integration is non-zero and zero means integration is zero for a given frequency. Thus, the origin of two SBs is due to non-zero value of $\int_{x1}^{x2} Image \times \cos(2\pi kx)\, dx$ for $k = Q$. As the hologram has a cosine fringe with some background and different higher amplitude than the FT cosine function, multiplication with another cosine function with same frequency match will yield high value of integration (see illustration in Fig. S9). That is why SB central spot intensity is symmetric to CB (see Fig. S10). Now the frequency $Q + ng$, where $g$ is the reciprocal lattice vector and $n$ = 1, 2, 3…., will also yield spots similar to spots around CB but with less numbers with higher frequencies as the oscillation of trigonometric functions are high compared to frequencies near CB, giving smaller value of integration.

From the illustration in Fig. S10 one can now see that the information on the wave functions given by FFT spots around the SBs, has contribution not only from the intensity of crystal periodicity but also from cosine periodicity in the recorded image. Same is true for FFT spots around CB. This is because the $f(x)$ or image function is same both for CB and SB while performing FT.

Now FT of the experimental image can be written down as

$$F.T.(I_{total\ HRholo})\ for\ all\ frequencies = F.T.(I_{total\ HRholo})\ for\ frequency\ g = 0\ and\ \pm ng$$
$$+F.T.(I_{total\ HRholo}) * \delta(q-Q)\ for\ frequency\ g = Q\ and\ Q \pm ng$$
$$+F.T.(I_{total\ HRholo}) * \delta(q+Q)\ for\ frequency\ g = -Q\ and\ -Q \pm ng$$

(75)

Only the absolute of the above Eq. 64 will reveal the FFT spots.

Now the total intensity of CB is at least one order of magnitude higher than two SBs together (Fig. S10). In case of HRTEM it is the same by considering all the FFT spots with respect to the direct spot at the center. Therefore, we have taken approximately 1% of the mean vacuum intensity as the intensity to assign to the image wave function [Eq. 9]. Now from the Eq. 9 after removing all the constant background terms and then a division by $2\sqrt{\alpha I_0}$ term retain only $\sqrt{\beta I_0}$ term associated with the trigonometric function that will give the amplitude of the object exit wave function (OEW). In case of off-axis holography it is given by $2a(x, y)$ divide over two for one of the SBs [Eq. 10]. Kindly note that the individual $A_0 \exp(i\phi_0)$ or $A_i \exp(i\phi_i)$ wave function components will not give any intensity pattern, only the phase change and associated overall amplitude $2A_0A_i$ associated with the cosine term of $\psi_{HRTEM}\psi_{HRTEM}^*$ can be known from the interference experiment.

### 2.2.1. 90° phase shift between diffracted wave and primary incident wave

This has origin in Fresnel-Huygens construction to describe the phase shift of diffracted wave with respect to primary wave.

However, we can understand the physical picture on the origin of 90° phase shift with the consideration of diffraction or interference geometry as given in Fig. S11. The first figure showing the example diffracted wave vector $k_d$ propagating at some angle $\theta$ with respect to primary wave vector $k_0$. The second illustration showing the formation of interference pattern when the diffracted and primary waves meet due to focusing on the image plane. The horizontal component of diffracted wave vectors will interfere according to off-axis holographic principle and produce interference pattern in the image plane. Whereas the *z* components will produce interference pattern which can be observed only by changing focus and monitoring intensity for a fixed location on the image plane. As the carrier frequency of the interference fringe depends on $k_d \cos(90 - \theta)$ and $k_d \sin(90 - \theta)$ on in plane and along z directions, a small angle will produce typically fringe wavelength of the order of 10 Å, for a wavelength of electron 2 pm and scattering angle 0.1 degree. The fringe wavelength along z direction will be ~ 2pm. Therefore, it will be difficult to see the pattern with focus change along z direction unless instrument focus settings have the required step resolution and only contribute to uniform background. Thus, the waves contributing to two different interference patterns have 90° angular separation and can be thought of as 90° phase shift between diffracted and primary waves and resulting flux distribution.

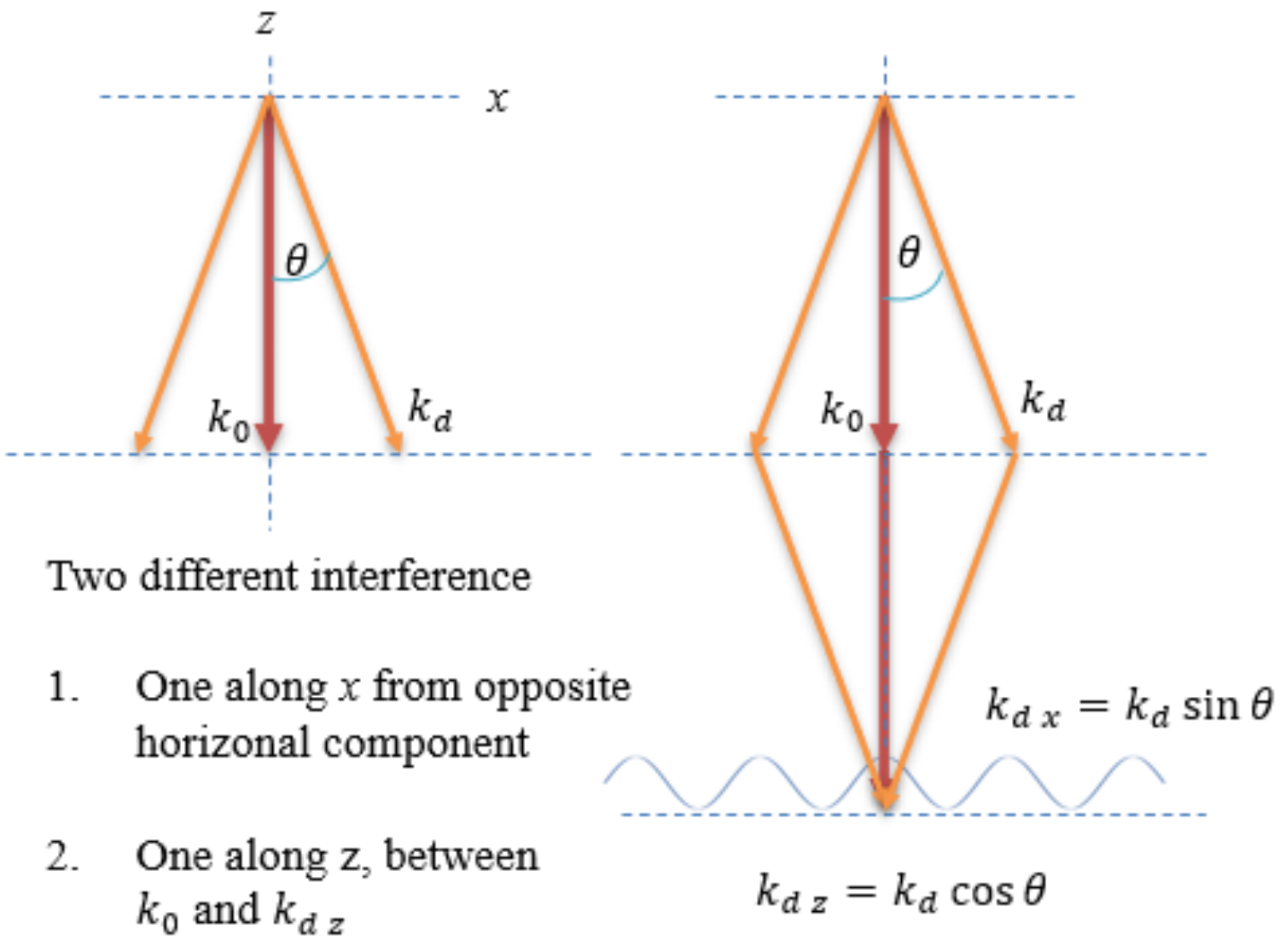

***Fig.S11.*** *Schematic representation of wave vector components taking part into interference. Horizontal components are responsible for intensity pattern while vertical components will give background and pattern corresponding to vertical components can only be observed with focus variation if focus steps of the order of wavelength of electron is available in an instrument.*

## 2.3. Paraboloid method of reconstruction

There are two different approaches discussed by Van Dyck and the second one by him has similarity with Kirkland method.

In the first approach of Van Dyck [16] what is widely known as 3D paraboloid method, the 2D Fourier transform $\psi(k)$of the wave function $\psi(x)$ is recovered via a summation

$$\psi(k) = \frac{1}{N}\sum_n I_n(k)\exp\{-\pi i\lambda z_n k^2\} \tag{76}$$

Based on large set of images recorded at equally and closely spaced focus level either side of the Gaussian focus. After summation over N images, the simplified restoring filter takes the form as

$$r_n(k) = \exp\{i\gamma_n(k)\}\frac{1}{N} \tag{77}$$

And the restored wave function becomes

$$\psi(k) = \sum_n I_n(k)\exp\{i\gamma_n(k)\} = \exp(\pi i C_s\lambda^3 k^4) \times \frac{1}{N}\sum_n I_n(k)\exp(-\pi i\lambda z_n k^2) \tag{78}$$

In the above expression, spherical aberration phase shift is included explicitly. For perfect coherence after appropriate approximation the restoring filter has the same form as in Eq. (54) above [based on Wiener formulation, see Schiske, Saxton and Hawkes].

What Saxton pointed out that in 3D paraboloid method, the picture can be obtained by 3D Fourier transformation of Eq. 6, i.e., to the third image space dimension *z* as well.

$$\delta(k,k_z) + \psi(k)\delta\left(k_z - \frac{1}{2}\lambda k^2\right) + \psi^*(-k)\delta\left(k_z + \frac{1}{2}\lambda k^2\right) + \int H(k,z)\exp(-2\pi i k_z z)\,dz \tag{79}$$

It was shown by Van Dyck that the wanted wave function $\psi(k)$ is localized on the paraboloid at $k_z = \frac{1}{2}\lambda k^2$ in reciprocal space and the complex conjugate will lie on a reflected paraboloid and completely separated except at small *k*.

At $k_z = \frac{1}{2}\lambda k^2$, the transformation with respect to *z* gives

$$\int I(k,z)\exp\{-\pi i\lambda k^2 z\}\,dz \quad (80)$$

Which is equivalent to Eq. S65.

### 2.4. Reconstruction based on partial coherence theory

In the second approach by Van Dyck *et al* ,[12] the method is based on partial coherence theory considering the effect of finite size of source, chromatic defocus spread, current voltage fluctuation of the instrument, objective aperture size and wave aberration function.[17,18] The reconstruction method based on partial coherence theory is known as iterative linear restoration which addresses the residual non-linear term. By repeated application of linear restoring filter from the subtraction of calculated non-linear term improves the initially guessed wave function.

According to Van Dyck, in frequency space the specimen to image interference is given by

$$I(G) = \phi(0)\phi^*(-G)T(0,-G) + \phi^*(0)\phi(G)T(G,0) + \int_{G'\neq 0, G'\neq G} \phi(G+G')\phi^*(G')T(G+G',G')dG' \quad (81)$$

Where G (G≠0) is the two-dimensional frequency vector. T is the transmission cross coefficient. The first and second term in Eq. S70 represents the linear interference term between the transmitted electron beam and one of the diffracted electron beams. The third term is the non-linear term involving interference between diffracted beams. It is the overlap between the two wave functions describes the correlation and contribute to the Fourier amplitude in the image. Thus, one can notice that the coherence and interference have origin in convolution operation.

✓ **Note on Eq. S70:** The above equation can be true is there is an integration outside the two wave functions term. The TCC can then be written equal to the convolution of two such wave functions within the integration. See the derivation for the same by Kirkland[4]

The same equation above has been reproduced by Saxton in his review under the section various other methods as follows,

$$I_n(k) = \delta(k) + \psi(k)w_n(k) + \psi^*(-k)w_n^*(-k) + \sum_{k'} \psi(k')\psi^*(k'-k)m_n(k', k'-k) \tag{82}$$

Where, $w(k)$ are the transfer functions and $m(k_1, k_2)$ is a mutual transfer function. The above equation is derived from Eq. (43) after considering the coherence effect.

The method described above also known as iterative linear restoration, addresses the residual non-linear term. By repeated application of linear restoring filter from the subtraction of calculated non-linear term improves the initially guessed wave function.

✓ **Note on Eq. S62:** In the equation above by Saxton, as the primary wave function is considered to be 1, thus one can avoid integration and convolution between wave function and transfer function becomes multiplication in reciprocal space. Kindly see below for unified description of above two equations

$$g(x) = \{(\psi_0 + \psi_i) \otimes h\}\{(\psi_0^* + \psi_i^*) \otimes h^*\}$$

$$G(k) = \{\psi_0(k)H(k) + \psi_i(k)H(k)\} \otimes \{\psi_0^*(k)H^*(k) + \psi_i^*(k)H^*(k)\}$$

$$= \int \{\psi_0(k')H(k') + \psi_i(k')H(k')\}\{\psi_0^*(k'+k)H^*(k'+k) + \psi_i^*(k'+k)H^*(k'+k)\}d^2k' \tag{83}$$

Without transfer function the above expression becomes

$$g(x) = \{(\psi_0 + \psi_i)\}\{(\psi_0^* + \psi_i^*)\}$$

$$G(k) = \{\psi_0(k) + \psi_i(k)\} \otimes \{\psi_0^*(k) + \psi_i^*(k)\}$$

$$= \int \{\psi_0(k') + \psi_i(k')\}\{\psi_0^*(k' + k) + \psi_i^*(k' + k)\}d^2k'$$

(84)

In Kirkland method, also called ML, MAP and MIMAP, attempts to solve the non-linear imaging problem exactly for the following equation set

$$I_n(k) = \sum_{k'} \psi\,(k')\psi^*(k' - k)m_n(k', k' - k) \qquad (85)$$

The above equation is re-written from Eq. S63 with the primary beam treated like other instead separately. As there is no obvious relationship with the Paraboloid method, a long focal series may not be required.

## 2.5. Partial coherence theory-based formulations used by others

The Fourier transformation of the above equation is written as according to Ref[1]

$$I(g) = \sum_k \psi(g + k)\psi^*(k) \qquad (86)$$

The above expression has been interpreted as pair wise summation of interference between two beams with wave vectors (*g*+*k*) and *k*. This is the description of wave function in the reciprocal plane which is the auto correlation between waves propagating along different direction governed by reciprocal lattice vectors.

The above result is due to the direct Fourier transformation of the image intensity in real space as following

$$I(r) = \psi(r)\psi^*(r) \tag{87}$$

According to Kirkland [p 99],[4] if one adds lens response

$$g(x) = |\psi_t(x) \otimes h_0(x)|^2 = [\psi_t(x) \otimes h_0(x)][\psi_t^*(x) \otimes h_0^*(x)] \tag{88}$$

The Fourier transform of the above equation leads to

$$G(k) = [\psi_t(k)H_0(k)] \otimes [\psi_t^*(k)H_0^*(k)]$$

$$= \int \psi_t(k')H_0(k')\psi_t^*(k'+k)H_0^*(k'+k)d^2k'$$

$$= \int T_{cc}(k',k'+k)\psi_t(k')\psi_t^*(k'+k)d^2k'$$

(89)

Where,

$$T_{cc}^{coh}(k',k'+k) = \exp\,[-i\chi(k') + i\chi(k'+k)]A(k')A(k'+k) \tag{90}$$

It is the overlap between two wave function describes the correlation and contribute to Fourier coefficient *k* in the image. Thus, one can notice that the coherence is because of convolution.

### 3. Note on Schiske filter function:

The purpose of Schiske's filter function was to carry out posterior correction of photograph records.[19] According to Schiske, the weakly scattered wave function $\psi_s(x)$ produced while reading the photographic records with incident wave $\psi_0$, which produces complex amplitude $a(x)$.

Scattered wave is written as

$$\psi_s(x) = a(x)\psi_0(x) = A_0 e^{i(k_0 . x_0)} \tag{91}$$

In Fourier space, $a(x)$ has the following form

$$a(x) = \int c(k) e^{i(k.x)} dk \tag{92}$$

Where, $c(k)$ is the amplitude of Fourier waves in frequency space and is written in terms of a complex function and its conjugate with the consideration of aberration function as

$$j(k) = c(k)\exp[-i\gamma(k_0 + k) + i\gamma(k_0)] + c^*(-k)\exp[i\gamma(k_0^* - k^*) - i\gamma(k_0^*)] \tag{93}$$

Finally, the Intensity in the image plane immediately beneath the object is written as

$$I(Mx) = I_0 + I_0 \int j(k)\, e^{i(k.x)} dk \tag{94}$$

Where, the first term in right hand side will give the central band or direct component, and the second term will form the scattered wave $\psi_s(x)$ and its conjugate. $\psi_s(x)$ is now known as OEW function in modern HRTEM literature.

Now, $c(k)$ is recovered by recording *n* images at various focus settings, and in simplified form by considering $k_0 = 0$,

$$c(k) = \frac{i}{2} \frac{\sum_m j_m(k) \sin\gamma_n}{\sum_n \sin^2\gamma_n} \tag{95}$$

Where, $\gamma_n$ is the aberration function with $n^{\text{th}}$ defocus.

Once, $c(k)$ is determined, $a(x)$ can be known. $a(x)$ is imaginary for pure phase object like weak phase object. And then the $\psi_s(x)$ can be determined without any residual aberration. This $\psi_s(x)$ has the information on the object structure.

What Schiske shown was to get rid of aberration from the information wanted, i.e. $\psi_s(x)$. But did not addressed the issue of twin image, and non-linear components as well as the amplitude of the scattered wave, which have contributions from conjugate and DC part.

## 4. Examples of real space fringe bending in off-axis electron holography

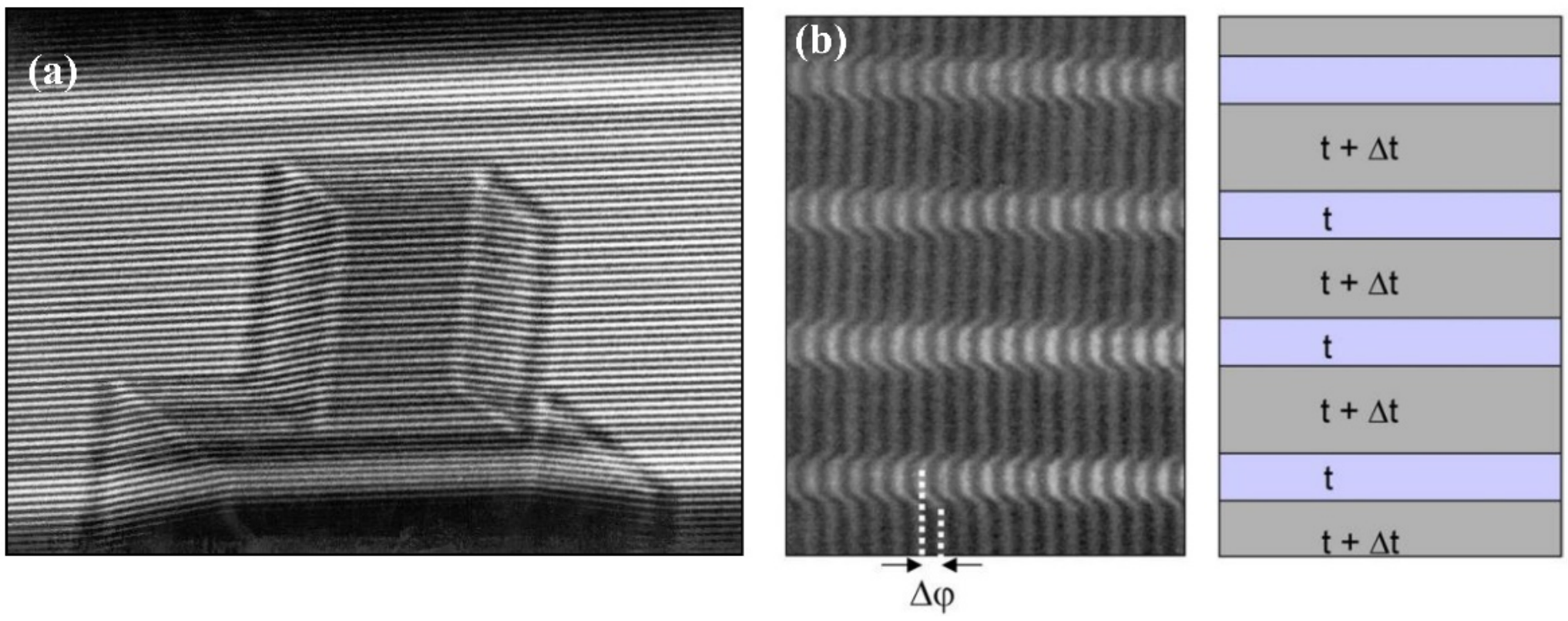

***Fig. S12.*** *Off-axis electron holography fringe bending after encountering object potential in image plane, (a) between vacuum and MgO crystal, and (b) object structured with different thickness stripes.*[14]

## 5. Evaluation of factors $\alpha$ and $\beta$

From the example micrograph in Fig. S13, one can determine the mean values of vacuum and mean intensity of image area where periodic dots are present with high low intensity. The standard deviation will be much smaller for vacuum area compared to image area. A typical example of vacuum mean intensity of MoS2 image is Mean=13133.8, standard deviation = 423.878, and image area mean is 13023.9 with standard deviation 1695.06. standard deviation in image area is approximately equal to $I_{\max dot}$ or $I_{\min dot}$. Now the values of $\alpha$ and $\beta$ are determined approximately by $\frac{I_{\max dot}}{I_{vacuum\ mean}}$ and $1 - \alpha$, respectively.

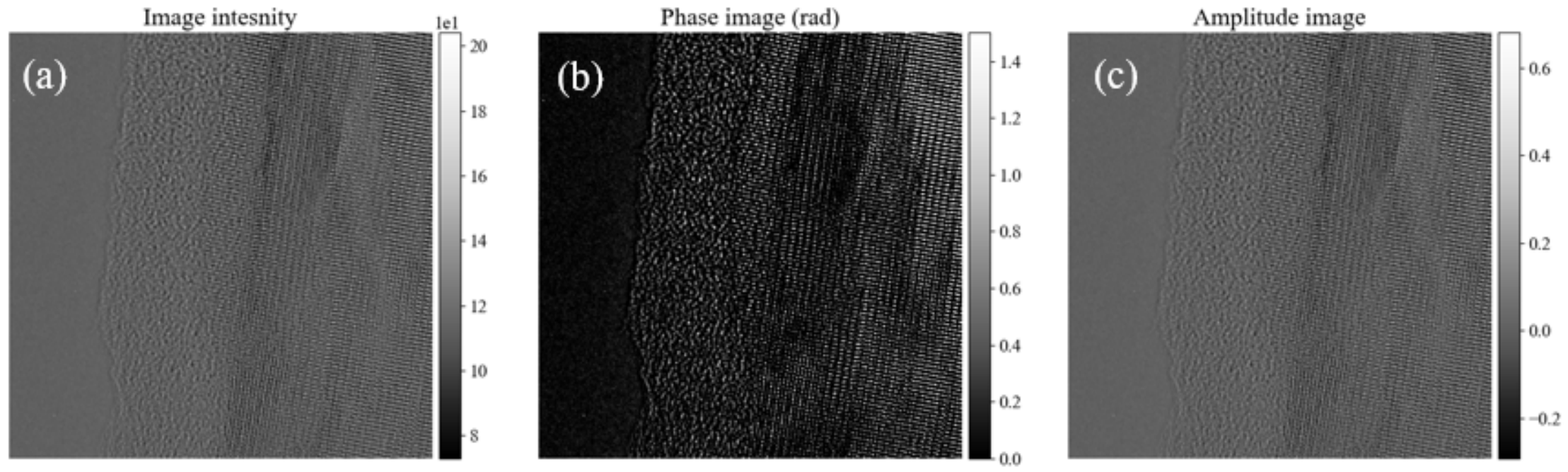

***Fig. S13.*** *(a) HRTEM image of ZnO epitaxial thin film monolayer recorded under negative* $C_S = -35\ \mu m$ *and positive defocus of* $\Delta f = 8$ *nm. (b) phase and (c) amplitude images of the OEW. Notice the systematic increase in phase and associated atom number for both Zn and O atoms away from the sample edge with the vacuum.*

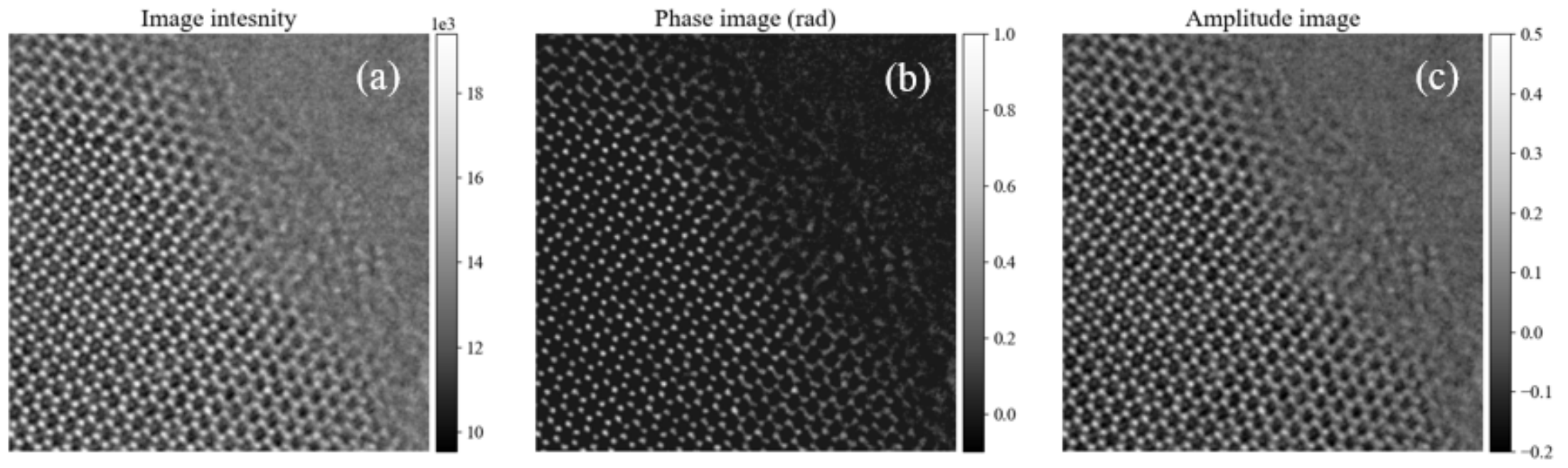

***Fig. S14.*** *Reconstructed phase and amplitude image of bilayer* $MoS_2$*.*

## 6. Peak phase shift values for different atoms

**Table S1.** Theoretical peak phase shift extracted for resolution in the range 1 - 0.5 Å,[20]

| Atoms (Z) | Phase shift (rad) |
|---|---|
| **B (5)** | 0.076-0.078 |
| **N (7)** | 0.100-0.102 |
| **O (8)** | 0.106-0.108 |
| **S (16)** | 0.164-0.23 |
| **Zn (30)** | 0.242-0.41 |
| **Mo (42)** | 0.300-0.51 |

## 6. Fourier series equations for retrieving image function from diffraction pattern

For isolated 2D function following Eq. 11 is modified to

$$f(x,y) = abs\left(\sum_{n=-k}^{n=k} \frac{1}{n} C_{k_x k_y} \cos\left(2\pi k_{k_x k_y} r\right)\right) \tag{96}$$

Where, $r = \sqrt{x^2 + y^2}$

However, for periodic 2D function one needs to modify the Eq. 11 to the following form including various spatial variables with cosine function.

$$f(x,y) = \sum_{n_{i,j}=k_{i,j}} \frac{1}{n_{i,j} \times n_{i,j}} C(k)_{i,j} \frac{1}{2} [\cos(2\pi k_{i,j} x) + \cos(2\pi k_{i,j} y)] + \frac{1}{(n_{i,j} \times n_{i,j}) \times (p \times q)} \frac{1}{4} C(k)_{i,j} [\cos(2\pi k_{i,j}(x+y)) + \cos(2\pi k_{i,j}(x-y))] \tag{97}$$

*Absolute* of right-hand side of Eq. S86 is not shown but need to consider obtaining a fit. One need to ensure to have more sampling points to have exact fit. The example for periodic Gaussian function in Fig. S19 used 400×400 sampling points.

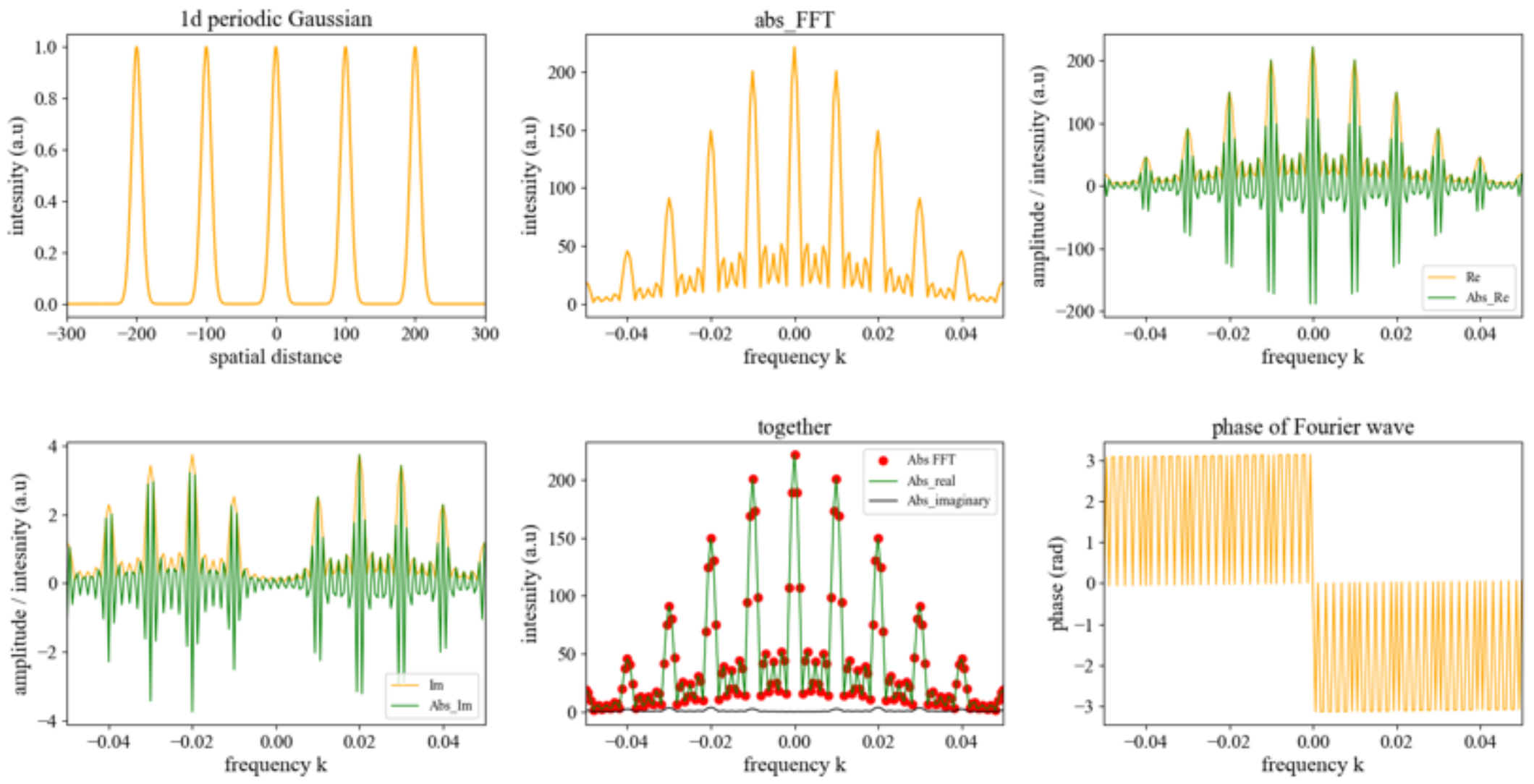

***Fig.S15.*** *(a) Example function, (b) Absolute FFT, absolute of (c) Re and (d) Imaginary parts. (e) contribution to absolute FFT is mostly from absolute Re part and contribution from absolute imaginary part is negligible.*

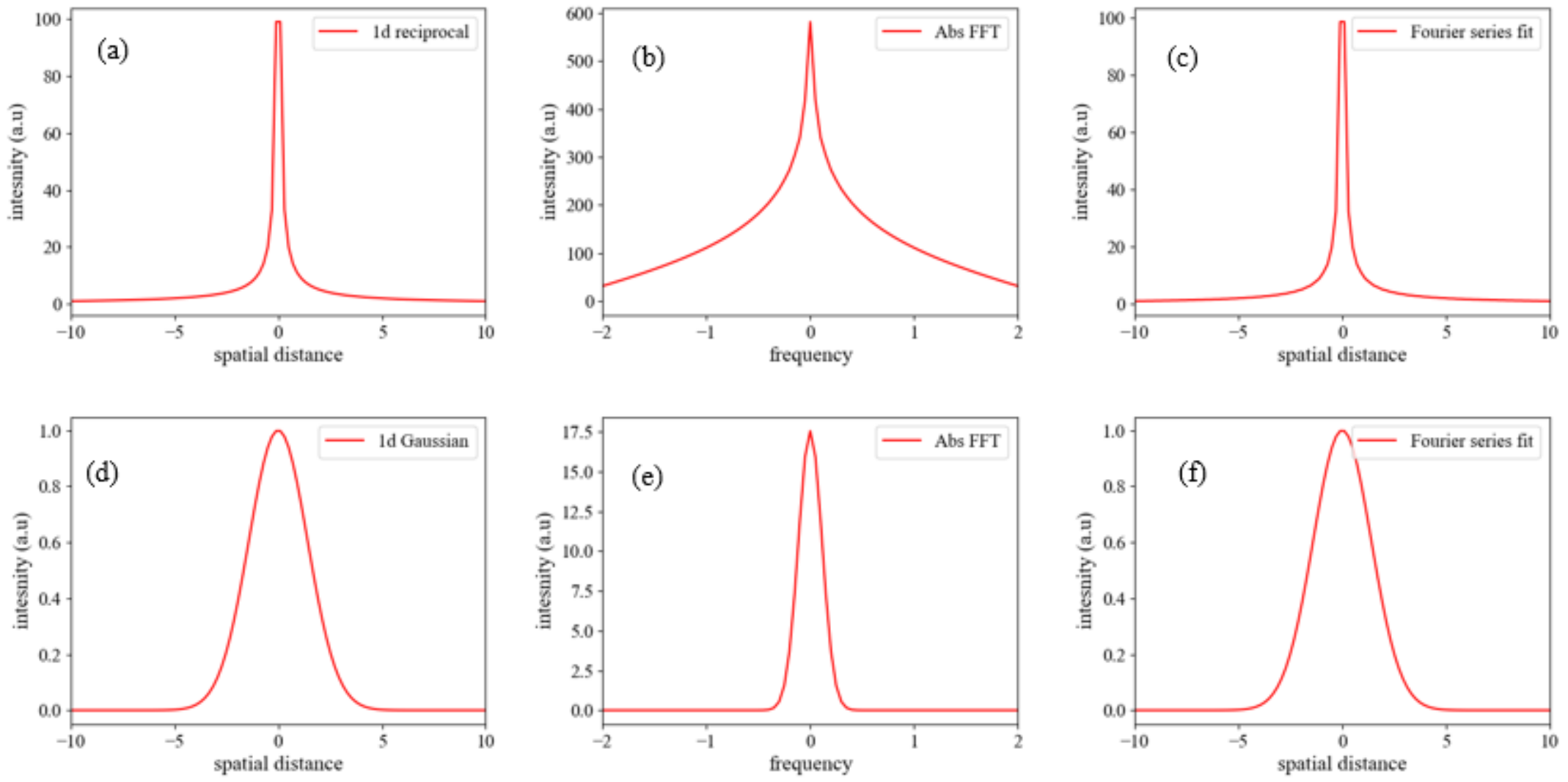

***Fig. S16****. example reconstruction for (a) reciprocal function, (b) Gaussian function in 1-dimensional form.*

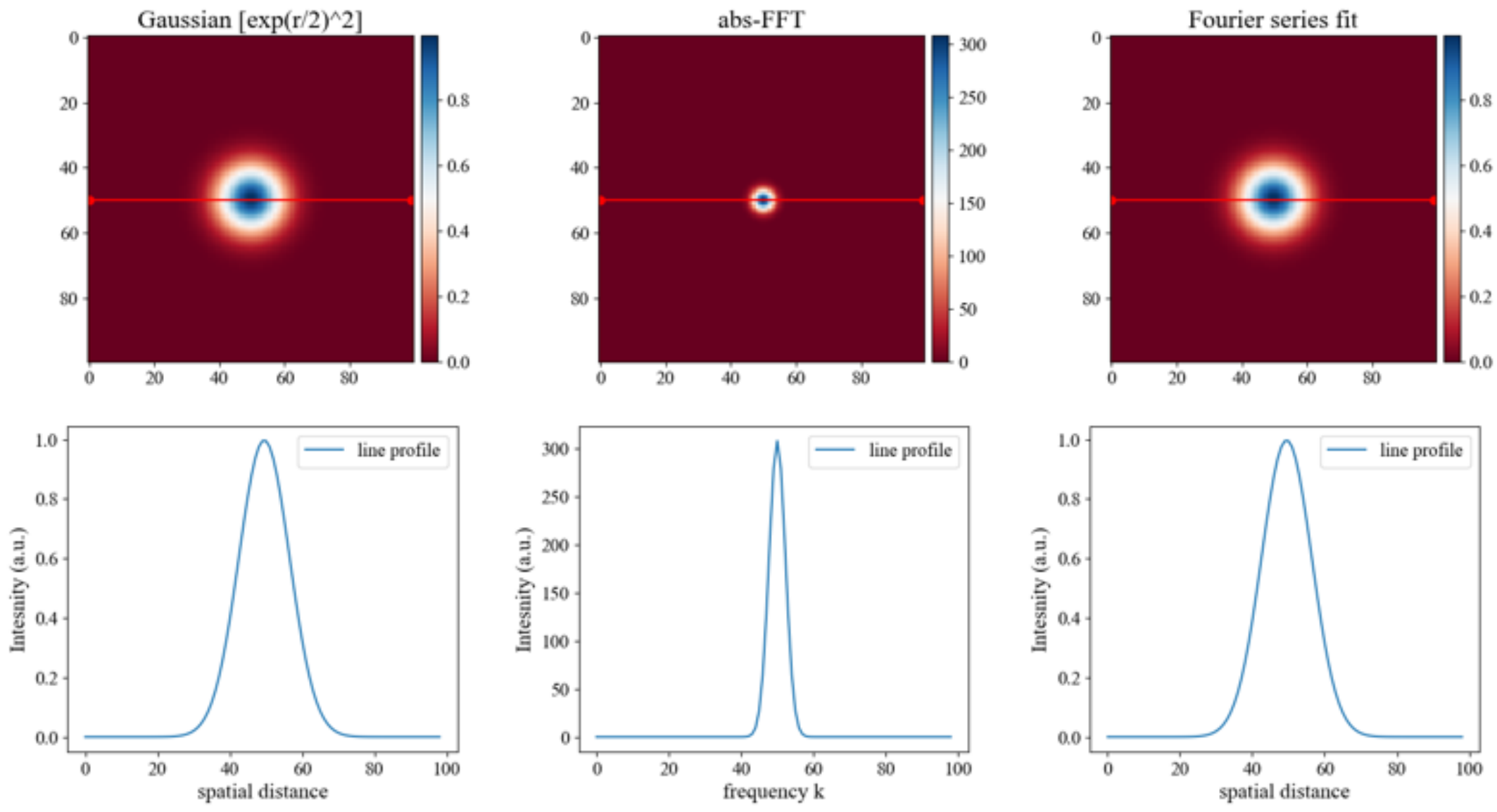

***Fig. S17****. example reconstruction for Gaussian function in 2-dimensional form with line profiles.*

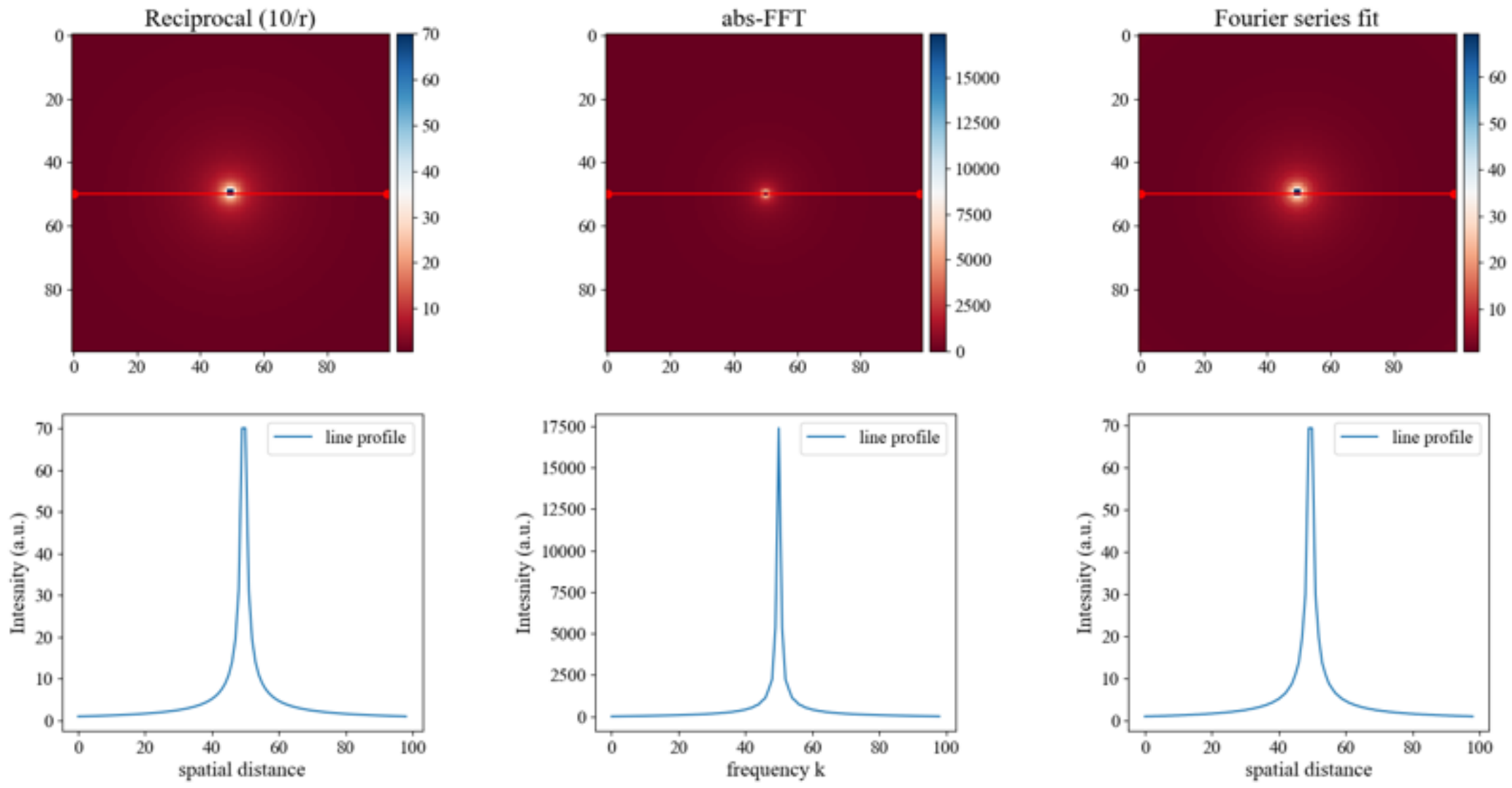

***Fig. S18****. example reconstruction for reciprocal function in 2-dimensional form with line profiles.*

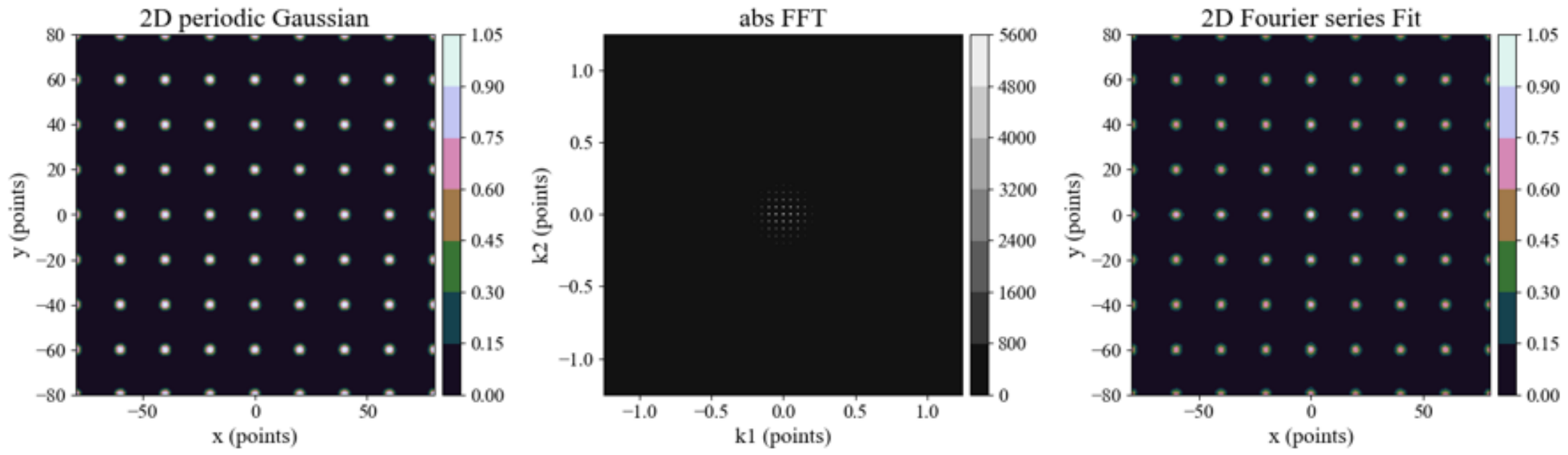

***Fig. S19***. *example reconstruction for periodic Gaussian function in 2-dimensional form.*